\documentclass[lettersize,journal]{IEEEtran}
\usepackage{algorithm2e}
\usepackage{algpseudocode}
\usepackage{graphicx,color}
\usepackage[normalem]{ulem}
\useunder{\uline}{\ul}{}
\usepackage{textcomp}
\usepackage{multirow}
\usepackage{hyperref}
\usepackage{cite}
\usepackage{amsmath,amssymb,amsfonts}
\usepackage{graphicx}
\usepackage[table]{xcolor}
\usepackage{colortbl}
\usepackage{tikz}
\usepackage{subfigure}
\usepackage{tabularx}
\usepackage{textcomp}
\def\BibTeX{{\rm B\kern-.05em{\sc i\kern-.025em b}\kern-.08em
    T\kern-.1667em\lower.7ex\hbox{E}\kern-.125emX}}

\usepackage{scalerel}
\usepackage{tikz}
\usetikzlibrary{svg.path}

\definecolor{orcidlogocol}{HTML}{A6CE39}
\tikzset{
  orcidlogo/.pic={
    \fill[orcidlogocol] svg{M256,128c0,70.7-57.3,128-128,128C57.3,256,0,198.7,0,128C0,57.3,57.3,0,128,0C198.7,0,256,57.3,256,128z};
    \fill[white] svg{M86.3,186.2H70.9V79.1h15.4v48.4V186.2z}
                 svg{M108.9,79.1h41.6c39.6,0,57,28.3,57,53.6c0,27.5-21.5,53.6-56.8,53.6h-41.8V79.1z M124.3,172.4h24.5c34.9,0,42.9-26.5,42.9-39.7c0-21.5-13.7-39.7-43.7-39.7h-23.7V172.4z}
                 svg{M88.7,56.8c0,5.5-4.5,10.1-10.1,10.1c-5.6,0-10.1-4.6-10.1-10.1c0-5.6,4.5-10.1,10.1-10.1C84.2,46.7,88.7,51.3,88.7,56.8z};
  }
}

\renewcommand{\arraystretch}{1.3} % Adjust table row spacing
\newcommand{\circled}[1]{\tikz[baseline=(char.base)]{
    \node[shape=circle,draw,inner sep=1pt] (char) {#1};}}
\newcommand\orcidicon[1]{\href{https://orcid.org/#1}{\mbox{\scalerel*{
\begin{tikzpicture}[yscale=-1,transform shape]
\pic{orcidlogo};
\end{tikzpicture}
}{|}}}}

\begin{document}

\title{A Threshold-Triggered Deep Q-Network-Based Framework for Self-Healing in Autonomic Software-Defined IIoT-Edge Networks}
\author{Agrippina Mwangi*\orcidicon{0000-0002-3663-8335}\,, León Navarro-Hilfiker \orcidicon{0009-0008-7375-4002}\,, Lukasz Brewka \orcidicon{0000-0000-0000-0000}\,, Mikkel Gryning \orcidicon{0000-0003-3653-4920}\,, Elena Fumagalli \orcidicon{0000-0002-5387-2860}\,, Madeleine Gibescu \orcidicon{0000-0002-4420-8538}\,
        % <-this % stops a space
\thanks{Manuscript received: xxxxxxxxxxx, revised: xxxxxxx, accepted: xxxxxxxxxx }
\thanks{The Innovative Tools for Cyber-Physical Energy Systems (InnoCyPES) project has received funding from the European Union’s Horizon 2020 research and innovation programme under the Marie Sklodowska Curie grant agreement No 956433.}

\thanks{Agrippina Mwangi, Elena Fumagalli, and Madeleine Gibescu are with the Energy and Resources Group at the Copernicus Institute of Sustainable Development, Utrecht University, The Netherlands. \\ (Corresponding Author Email: a.w.mwangi@uu.nl)}

\thanks{León Navarro-Hilfiker and Lukasz Brewka are with the Operational Technology (OT) System Engineering and Security (Engineering, Procurement, and Construction) at Ørsted Wind Power.} 
\thanks{Mikkel Gryning is with the System Design Specialists (Engineering, Procurement, and Construction) at Ørsted Wind Power.} 

}

\markboth{IEEE TRANSACTIONS ON NETWORK AND SERVICE MANAGEMENT, \LaTeX\ Class Files,~Vol.~, No.~, April~2025}%
{Mwangi \MakeLowercase{\textit{et al.}}: A Threshold-Triggered Deep Q-Network For Self-Healing in Autonomic Software-Defined IIoT-Edge Networks}

\IEEEpubid{0000--0000/00\$00.00~\copyright~2025 IEEE}

\maketitle

\begin{abstract}
%%%%%%%%%% PROBLEM STATEMENT
Stochastic disruptions such as flash events arising from benign traffic bursts and switch thermal fluctuations are major contributors to intermittent service degradation in software-defined industrial networks. These events violate IEC~61850-derived quality of service requirements and user-defined service-level agreements, hindering the reliable and timely delivery of control, monitoring, and best-effort traffic in IEC~61400-25-compliant wind power plants. Failure to maintain these requirements often results in delayed or lost control signals, reduced operational efficiency, and increased risk of wind turbine generator downtime.
%%%%%%%%%% SOLUTIONS
To address these challenges, this study proposes a threshold-triggered Deep Q-Network self-healing agent that autonomically detects, analyzes, and mitigates network disruptions while adapting routing behavior and resource allocation in real time. The proposed agent was trained, validated, and tested on an emulated tri-clustered switch network deployed in a cloud-based proof-of-concept testbed.
%%%%%%%%%% RESULTS
Simulation results show that the proposed agent improves disruption recovery performance by 53.84\% compared to a baseline shortest-path and load-balanced routing approach, and outperforms state-of-the-art methods, including the Adaptive Network based Fuzzy Inference System by 13.1\% and the Deep Q-Network and Traffic Prediction-based Routing Optimization method by 21.5\%, in a super-spine leaf data-plane architecture.
%%%%%%%%%% DISCUSSION AND CALL FOR ACTION
Additionally, the agent maintains switch thermal stability by proactively initiating external rack cooling when required. These findings highlight the potential of deep reinforcement learning in building resilience in software-defined industrial networks deployed in mission-critical, time-sensitive application scenarios.
\end{abstract}

%These findings underscore the potential of deep learning algorithms to enhance software-defined networks by fostering resilience and autonomy in critical industrial operational technology systems.


%Self-healing, DQN, ETSI, Hybrid-band control, IIoT-Edge, IEEE802.1 TSN, IEC60870-5, IEC62439-4/5, IEC61850, IEC60309-1/2, Mininet, NFV, Offshore wind, ONOS, Resilience, SDN, SLA, QoS.

%\vspace{1.5em}
%%%%%%%%%%%%%%%%%%%%%%%%%%%%%%%%%%%%%%%%%%%%%%%%%%%%%%%%
\section{INTRODUCTION}
%%%%%%%%%BACKGROUND AND CONTEXT%%%%%%%%%%%%
Offshore wind power plant (WPP) operators are set to adopt vendor-agnostic software-defined networking (SDN) and network function virtualization (NFV) technologies to create scalable, programmable, and centrally managed industrial networks \cite{hutton2021deploying}. These SDN/NFV solutions simplify the management of large-scale, heterogeneous industrial networks, accelerating the deployment of new services and innovation to support continuous, real-time data exchange for critical, time-sensitive, and best-effort data traffic \cite{mwangi2024towards}. 
Further, WPP operators, in compliance with international standards such as IEC60870-5, IEC61850-5, and IEC62439-4/5, enforce stringent Service Level Agreements (SLA) and Quality of Service (QoS) requirements to ensure highly available and highly performing industrial networks, guaranteeing real-time WPP operations and associated service provision \cite{vizarreta2019incentives}. 

%Orchestration of these Industry 4.0 technologies hinges on a robust communication infrastructure. https://www.energy.gov/sites/default/files/2024-05/Understanding%20and%20Managing%20Quality-of-Service%20in%20Grid%20Communications.pdf

%%%%PROBLEM STATEMENT %%%%%%
While SDN/NFV-based solutions present myriad benefits, this study identifies two main challenges that cause intermittent network service interruptions: 

\textbf{\circled{A}} Flash events of benign traffic flows that often occur as a result of regular offshore WPP events such as (i) state changes in wind turbines, which trigger bursts of data exchange between the centralized protection and control system and the actuators, (ii) SCADA polling during maintenance checks or diagnostic cycles, and (iii) scheduled firmware or software updates \cite{mwangi2023building}. These brief bursts of benign traffic cause network congestion, leading to intermittent service interruptions \cite{kanagavelu2019pro}. This directly affects the timeliness, reliability, and security of WPP industrial control and remote monitoring systems and, in extreme cases, often leads to unplanned wind turbine shutdowns.  
\IEEEpubidadjcol

\textbf{\circled{B}} Additionally, the prolonged processing of large volumes of network traffic, together with the variability of offshore thermal conditions, produces fluctuating temperature profiles across Ethernet switch devices \cite{buyya2010energy}. Such fluctuations are often harmful: elevated temperatures accelerate material degradation, reducing device lifespan by as much as 50\% and increasing error rates and packet drops \cite{ASHRAE_TC0909_2016}, while excessively low temperatures can make materials brittle and prone to physical failure. 
In some cases, extreme temperature profiles trace back to faults or degraded performance in the Heating, Ventilation, and Air Conditioning (HVAC) units or rack-cooling fans within the WPP’s offshore digital substation. These issues may arise from ruptured ventilation pipes, exposure to high wind gusts, corrosion, or seismic disturbances \cite{mcquiston2023heating}. When operating correctly, the rack-fan systems (supported by the HVAC units) maintain stable temperature conditions consistent with IEC 60309-1/2 and IEC 60320, thereby preserving the overall integrity of the network equipment.

\begin{figure}
    \centering
    \includegraphics[width=\columnwidth]{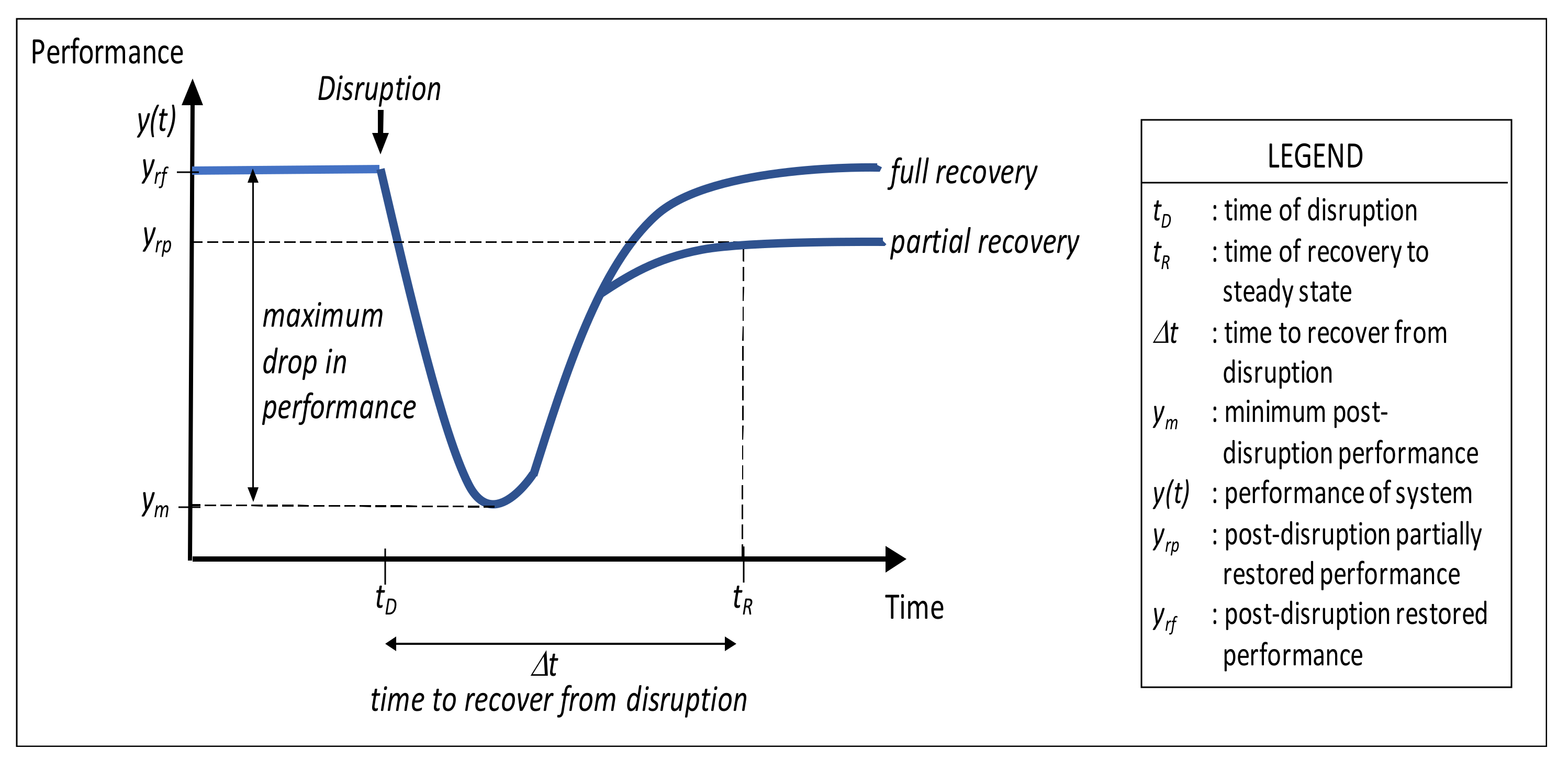}
    \caption{A time-based performance curve for a software-defined industrial IIoT-Edge network before, during, and after disruption, within time [0,t) adapted from \cite{madni2020constructing}.}
    \label{fig:res_curve}
\end{figure}

Figure \ref{fig:res_curve} shows a time-based network performance curve, conceptualized by \cite{madni2020constructing}, depicting the network performance shifts before, during, and after stochastic disruptions \textbf{\circled{A}} and \textbf{\circled{B}}.
%(i.e., flash events of benign traffic flows and deviations in IIoT-Edge network equipment thermal state from established nominal temperature profiles).
When a disruption occurs at time $t_D$, the network performance drops to the minimum post-disruption level, $y_m$. 
For an offshore WPP operator, such a drop in performance may result in several negative consequences: \textbf{\circled{i}} loss of critical data related to several crucial WPP data services which cause delayed response and actuation, \textbf{\circled{ii}} increased operational expenses (OPEX) from network maintenance costs and emergency repairs, and ultimately, and \textbf{\circled{iii}} revenue loss from unplanned WPP downtime and penalties for missed energy deliveries that may occur when WPP operators cannot efficiently manage and control the WPP components due to a lack of accurate, real-time data.

%%%%%% Network Performance Curve

To mitigate the aforementioned stochastic disruptions, novel approaches are adopted to minimize \textbf{\circled{i}} the performance drop  ($\Delta y = y_{rf} - y_{m}$) and \textbf{\circled{ii}} recovery time ($\Delta t = t_R - t_D$), aiming to swiftly restore the network performance either partially to ($y_{rp}$) or fully to  ($y_{rf}$) \cite{madni2020constructing,fonseca2017survey}. These novel approaches apply essential traffic engineering actions such as optimally allocating network resources, optimizing routing paths, prioritizing traffic through custom policies \cite{agarwal2013}, and distributing traffic loads evenly across the best paths 
\cite{boutaba2018comprehensive,rafique2018machine}. 
Due to the unpredictable nature of these disruptions, a self-healing process is required to restore this software-defined industrial network to optimal performance.

The European Telecommunications Standards Institute (ETSI) self-X (\textit{self-configure, self-heal, self-optimize, self-protect}) network and service management framework \cite{etsi-tr103747} aims to transform conventional industrial networks into cognitive, autonomic industrial networks. This paper focuses on self-healing, a crucial component of this framework.
\textit{Self-healing} is a reactive and autonomic process that detects, analyzes, and repairs malfunctions while adapting the network's behavior in response to stochastic disruptions \cite{liang2023self}. 
Self-healing agents are deployed through external adaptation or integrated mechanisms \cite{al2011autonomic}. The integrated self-healing agents are embedded in the control plane, enabling faster, more efficient self-healing. 
This approach is preferred for smaller, controlled environments, despite the higher initial costs and design effort. 
It is challenging to scale and adapt these integrated self-healing agents to other networks that use different types of SDN controllers (SDNC). 
Conversely, the externally adapted agents run on an additional layer (application or knowledge plane) and interact with the network's control plane via the Representational State Transfer Application Programming Interface (RESTful API). 
This makes them ideal for dynamic, large-scale systems where direct integration is impractical or costly, as is the case for the large-scale offshore WPPs' communication networks. 

%%%%%%%%%%%%%%%%%%%%%%%    RELATED WORKS  AND METHODOLOGIES PREVIOUSLY USED %%%%%%%%%%%%%%%%%%%%%%%%%%%%%%%%
Empirical studies have demonstrated the use of deep reinforcement learning (DRL) for both integrated and externally adapted self‑healing agents in diverse SDN environments.  Deep Deterministic Policy Gradient (DDPG) and Deep Q-Network (DQN)- based integrated agents have been successfully evaluated in smart city infrastructure, Internet backbone segments, and vehicular ad hoc networks \cite{dake2021multi,casas2020intelligent,casas2021drsir}. Similarly, externally adapted DRL agents have proven their effectiveness in large‑scale, dynamic settings where direct control plane integration is impractical by leveraging RESTful interfaces to effect autonomic recovery actions \cite{bouzidi2021deep,zhou2022aqrom,drom2018}. While effective, these approaches primarily focus on failure modes such as link outages or network congestion in environments with relatively stable Ethernet switch thermal loads and latency-tolerant service demands. This limits their applicability in offshore environments where thermal considerations are critical.

To address this gap, this paper proposes an externally adapted, threshold‑triggered DQN self‑healing agent designed for large‑scale, heterogeneous offshore WPP industrial networks.  Our proposed self-healing agent continuously monitors both traffic flow metrics and Ethernet switch temperature profiles, dynamically adjusting decision thresholds to distinguish benign surges from critical degradations.  By embedding these adaptive thresholds into a violation checker, the proposed agent achieves  \textbf{\circled{i}} reduced false‑positive interventions during harmless flash events, and  \textbf{\circled{ii}} accelerated corrective actions under genuine thermal or load‑induced stress.

%%%%%%%%%%%%%%%% Paper Contributions  %%%%%%%%%%%%%%%%%%%%%%%%%% 
\subsection{Paper contributions}
The key scientific contributions of this paper are:
\begin{enumerate}
    \item An autonomic, threshold-triggered Deep Q-Network self-healing agent (\texttt{TTDQSHA}) which is formulated as a Markov Decision Process problem that uses Deep Q-Networks to determine the best course of action when network traffic metric and Ethernet switch thermal metrics predefined thresholds or user intents are violated (see Section \ref{model}).
    \item A scalable cloud-based network emulator testbed \cite{ttdqshagit2025} integrates Mininet-modeled WPP switch fabrics with a three-node SDN controller cluster and the proposed self-healing agent via a northbound RESTful API, enabling realistic validation of autonomic recovery strategies (see Section \ref{part3}).
    \item A novel public dataset for offshore WPP software-defined networks, featuring traffic traces and thermal-performance metrics, to benchmark thermal-aware DRL-based self-healing in mission-critical environments \cite{mwangi2025dataset}.
\end{enumerate}

%%%%%%%%%%%%%%%%%%%% NOMENCLATURE %%%%%%%%%%%%%%%%%
\subsection{Nomenclature and Mathematical Notation}
This section presents the nomenclature and mathematical notation used in the proposed \texttt{TTDQSHA} framework for autonomic software-defined industrial networks in clustered offshore WPPs, organized in the order of appearance.

%\section*{nomenclature (in order of appearance)}
\begin{table}[!htbp]
\caption{NOMENCLATURE AND MATHEMATICAL NOTATION \\(IN ORDER OF APPEARANCE)}
\resizebox{\columnwidth}{!}{%
\begin{tabular}{ll}
\\
\hline
$\mathcal{G}$     & Non-oriented graph               \\ \hline
$\mathcal{V}$     & Set of network nodes (OpenFlow switches)             \\ \hline
$\mathcal{E}$     & Set of network links (switch-to-switch physical links)              \\ \hline
$\mathcal{F}$     & Set of flow rules              \\ \hline
$TM_t$          &   Traffic matrix             \\ \hline
$u_{i,t}$       &  Link utilization of link, $i$, at time, $t$             \\ \hline
$l_{j,t}$     &    Aggregated latency of source-destination path, $j$, at time, $t$           \\ \hline
$\tau_{k,t}$  & Aggregated switch temperature of switch, $k$, at time, $t$ \\ \hline
$s_t$  &  Network state \\ \hline
$\eta$  &   real-valued dimensional coordinate space equivalent to (n+k+m) \\ \hline
$a_t$  &  Action space \\ \hline
$p_{w,t}$   & Pre-configured redundant path, $w$, at time, $t$  \\ \hline
$f_{h,t}$   & Flows based on service type classification, $h$, at time, $t$ \\ \hline
$\mathcal{R}(s_t,a_t)$  &  Reward function \\ \hline
$\alpha, \beta$  &  Tunable parameters for the reward function \\ \hline
$\mathcal{C}$  &  Normalized network link bandwidth \\ \hline
$N_{req}$  & QoS/SLA thresholds from the Abstraction module \\ \hline
$u_{thr}$  & Link utilization threshold \\ \hline
$l_{thr}$  &  End-to-end path latency threshold \\ \hline
$\tau_{thr}$  &   Device temperature threshold \\ \hline
$\tau_{k}^{(ambient)}(t)$ & Ambient inlet temperature for switch, $k$, at time, $t$  \\ \hline
$\tau_{k}^{(internal)}(t)$ & Internal temperature profile for switch, $k$, at time, $t$ \\ \hline
$\mathcal{Q(\theta)}$  &  Neural network with random weight $\theta$ \\ \hline 
$\hat{\mathcal{Q}}\hat{(\theta)}$  &  Target neural network with random weight $\hat{\theta}$ \\ \hline 
$\mathcal{D}$  &  Experience replay memory \\ \hline
$\mathcal{M}$  & Number of training episodes  \\ \hline
$\gamma$ & Discount factor \\ \hline
$\epsilon$  & Exploration rate \\ \hline
$\xi$  &  \text{Uniform}(0, 1) \\ \hline
$\mathcal{U}_{k}(t)$  & Switch Utilization switch $k$ at time $t$\\ \hline
\end{tabular}%
}
\end{table}

%%%%%%%%%%%%%%% Organization of the paper %%%%%%%%%%%%%%%%%%%%%%%%%%%%%%%
\subsection{Organization of the paper}
The rest of the paper is organized as follows:
Section \ref{part2} outlines the background, reviews related work, and presents the autonomic software-defined IIoT-Edge communication networks in a clustered offshore WPP application scenario along with key design constraints.  
Section \ref{model} describes the threshold-triggered Deep Q-Network self-healing system model. 
The methodology in section \ref{part3} describes the design of the cloud-based proof-of-concept testbed, comprising Mininet-emulated super-spine-leaf switch network topologies at the data plane, ONOS-based controller clusters at the control plane, and a threshold-triggered DQN self-healing agent in the knowledge plane.
Section \ref{part4} highlights and discusses the key findings, industrial application recommendations, and insights into future work. Finally, section \ref{part5} concludes the paper.

\section{BACKGROUND AND RELATED WORK} \label{part2}
\begin{table*}[ht]
\centering
\caption{SUMMARY OF DEEP REINFORCEMENT LEARNING APPROACHES FOR SELF-HEALING IN SOFTWARE-DEFINED NETWORKS}
\label{tab:drL_sdn_routing}
\resizebox{\textwidth}{!}{%
\begin{tabular}{|l|p{3cm}|p{3cm}|p{3cm}|p{2.5cm}|p{3cm}|p{2.5cm}|}
\hline
\textbf{Study} & \textbf{Approach} & \textbf{Testbed Design} & \textbf{Key Mechanism} & \textbf{Evaluation} & \textbf{Effectiveness} & \textbf{Domain} \\
\hline
\cite{dake2021multi} & Multi-Agent DDPG (MA-DDPG) & RYU SDN controller, smart city IoT simulation with zombie devices & Collaborative policy learning among agents for routing and security & Simulation under varied IoT attack scenarios & Improved QoS/QoE; optimized routing and DDoS mitigation & SDN-IoT networks, smart city \\
\hline
\cite{casas2020intelligent} & Deep Q‑Network (DQN) & RYU SDN controller, intelligent routing framework & Discrete routing decisions & Dijkstra shortest-path & Outperformed Dijkstra in stretch, packet loss, and delay metrics & SDN routing optimization \\
\hline
\cite{casas2021drsir} & DQN within DRSIR framework & RYU + Mininet emulation & Discrete path-routing & Dijkstra, RSIR & Achieved superior adaptability to traffic dynamics, improved QoS & SDN routing optimization \\
\hline
\cite{bouzidi2021deep} & DQN + Traffic Prediction-based Routing Optimization (DPTRO) & Network flow routing with external DQN agent & Combines traffic prediction with reinforcement learning routing & Sensitivity and training complexity analysis & Dynamic and efficient routing, less sensitive than DDPG & SDN flow routing optimization \\
\hline
\cite{zhou2022aqrom} & Asynchronous Advantage Actor-Critic (AQROM) & OMNET++ network simulation; general load balancing scenario & QoS-aware routing with asynchronous advantage actor-critic & Load balancing metrics under different network loads & Efficient load balancing with QoS guarantees & General SDN routing, load balancing \\
\hline
\cite{drom2018} & DDPG Routing Optimization Mechanism (DROM) & OMNET++ simulation; Sprint backbone topology & Reinforcement learning-based adaptive routing policy & Comparative analysis with OSPF routing & Superior routing performance in delay and packet loss & Backbone network routing \\
\hline
\cite{hu2020ears} & DDPG automatic routing agent & POX SDN controller; ChinaNET topology (EARS experimental network) & Continuous action space routing optimization via DDPG & Performance metrics compared with OSPF, ECMP & Outperformed baseline algorithms in latency and throughput & SDN routing, experimental network \\
\hline
\cite{zhang2022} & Deep Q-learning approach & Software-defined trust-based vehicular ad-hoc network testbed & Trust-aware link quality optimization via DQN & Link quality and network stability evaluation & Enhanced link quality in vehicular networks & Vehicular ad-hoc networks (VANETs) \\
\hline
\end{tabular}%
}
\end{table*}

Machine Learning, with an emphasis on Reinforcement Learning (RL), has been instrumental in the design of self-healing agents for a broad range of application scenarios, including cyber-physical energy systems, control systems, and, more recently, communication networks \cite{luong2019applications,cao2020,latah2019artificial}. 
Sutton et al. \cite{sutton2018reinforcement} define RL as \textit{``a learning process in which an agent can periodically make decisions, observe the results, and then automatically adjust its strategy to achieve the optimal policy''}. 
RL-based self-healing agents are adept at learning sequential decision-making to apply traffic engineering actions at the control plane of the software-defined IIoT-Edge network, ultimately improving the network performance. 
Unfortunately, conventional RL-based agents require extensive training time to explore their environments and optimize their policies, making them impractical for large-scale communication networks. 
To address this, deep neural networks (DNNs) are integrated with RL principles to enable neural network entities to learn about and react to their 
environment, solve complex optimization problems, make autonomous decisions, and enhance the performance and speed of RL algorithms. The resultant approach, DRL, is a more viable alternative for complex, real-time decision-making, e.g., in communication networks \cite{dake2021traffic}.

\subsection{Deep Reinforcement Learning Approaches}
In their research, \cite{dake2021multi,hu2020ears,drom2018,zhou2022aqrom} introduce DRL-based self-healing and routing-optimization agents integrated into various SDN environments, demonstrating improved QoS and routing performance across smart-city, experimental, and backbone network scenarios.
Although these DDPG-based approaches have effectively addressed the traffic engineering control problem, they argue that DDPG, employing an actor-critic framework with deep neural networks, tends to converge slowly in large-scale settings with extensive state and action spaces. Moreover, adapting trained DRL agents to operate across different communication network environments has proven quite challenging.

As an alternative to the DDPG family of methods, empirical studies \cite{casas2021drsir,casas2020intelligent,zhang2022,bouzidi2021deep} explored DQN approaches for intelligent routing and traffic engineering in SDN and vehicular networks, demonstrating improved performance over traditional routing algorithms and enabling the adaptation of external agent architectures. These studies framed their action spaces as discrete choices, allowing the DQN to be highly effective. Furthermore, compared with DDPG, DQN agents are easier to train and less sensitive to parameter settings that require careful tuning. Table \ref{tab:drL_sdn_routing} summarizes these two DRL approaches, highlighting their technical methodologies, testbed configurations, evaluation metrics, and overall effectiveness across different SDN domains. Given the robust performance of DQN self-healing agents in environments with discrete action spaces, this paper considers DQN as a suitable approach for designing the self-healing framework.

%%%%% OFFSHORE WIND FARM SCENARIO%%%%%%%%%
\subsection{Application Scenario and Design Constraints}
%%%%%% Hornsea WPP cluster

\begin{figure}[!htbp]
    \centering
    \includegraphics[width=\columnwidth]{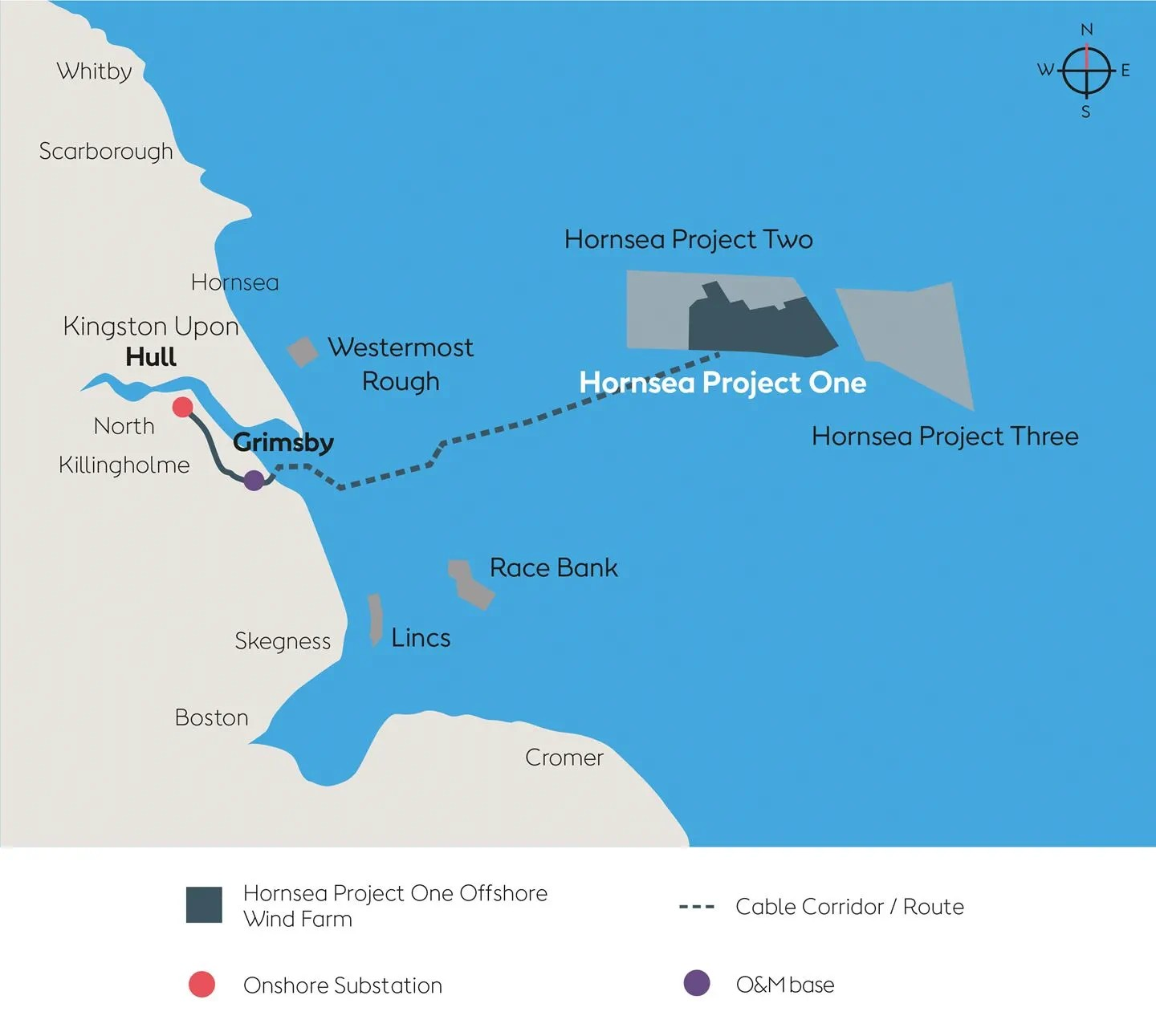}
    \caption{A schematic representation of the Hornsea offshore Wind Power Plant (WPP) cluster located $\geq100$ km off the Yorkshire coast in the North Sea. (Source: Ørsted UK, 2019).}
    \label{fig:hornsea}
\end{figure}

This study considers a clustered offshore WPP architecture, such as the Hornsea WPP cluster illustrated in Figure \ref{fig:hornsea}, in which each plant comprises multiple wind turbine generators (WTGs) connected to an offshore substation (OSS) reconfigured as a modular edge data center. This clustered arrangement offers operational resilience and system-wide flexibility by: \textbf{\circled{i}} enabling inter-OSS coordination to support substation-level failover and recovery, \textbf{\circled{ii}} allowing neighboring reactive compensation substations to mitigate power generation losses through dynamic reserve sharing, and \textbf{\circled{iii}} supporting predictive maintenance and dynamic load balancing across the cluster via wide-area monitoring and control.
%%%%%% Offshore Wind Farm
\begin{figure*}
    \centering
    \includegraphics[width=\textwidth]{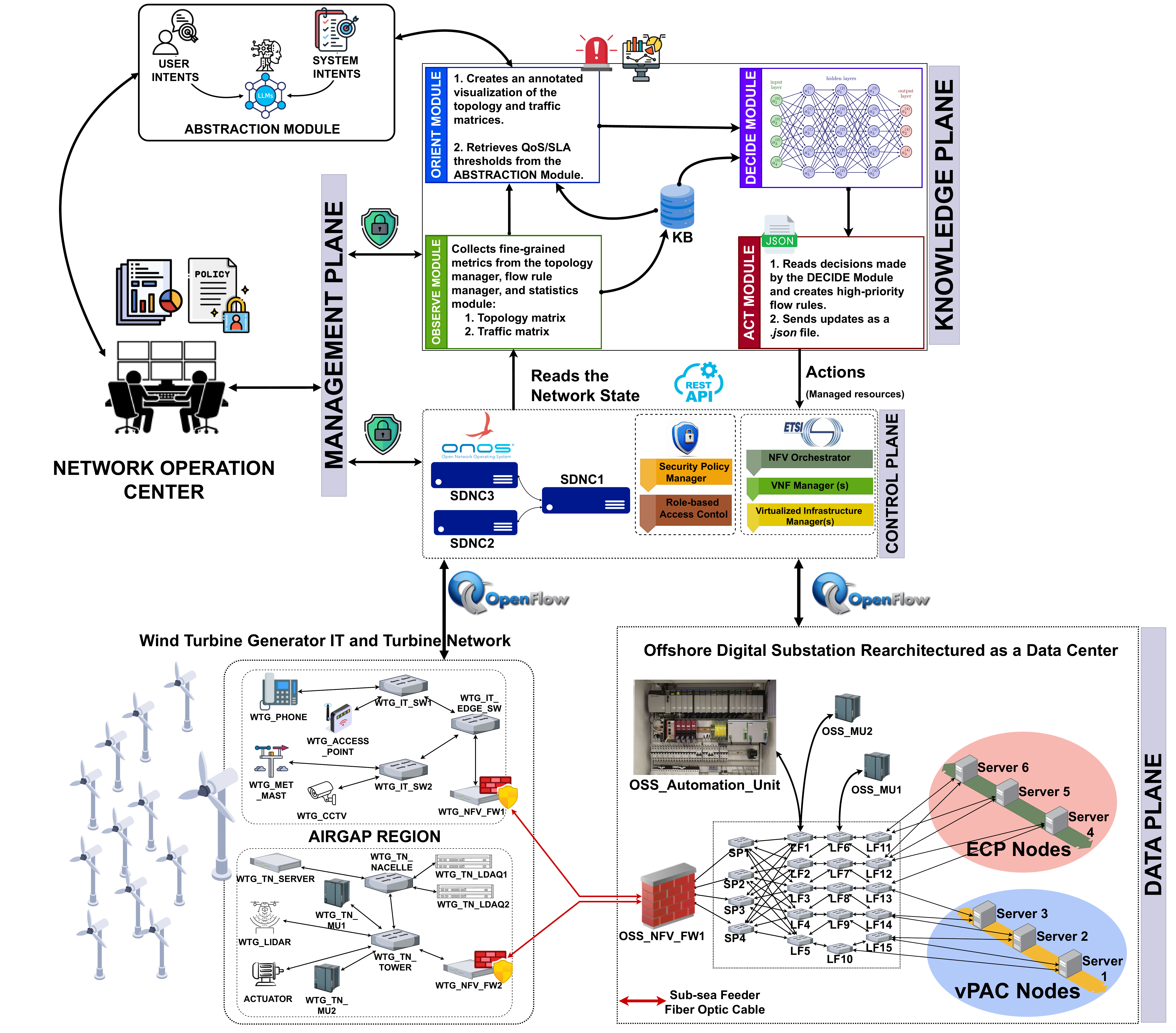}
    \caption{A hybrid-band control (HBC) software-defined operational IIoT-Edge network using OpenFlow protocol for a fleet of Wind Turbine Generators communicating with an offshore digital substation (OSS) re-architectured as a data center using fiber optic cables.}
    \label{fig:wpp}
\end{figure*}

%%%%%% Server Cluster
\begin{figure}
    \centering
    \includegraphics[width=\columnwidth]{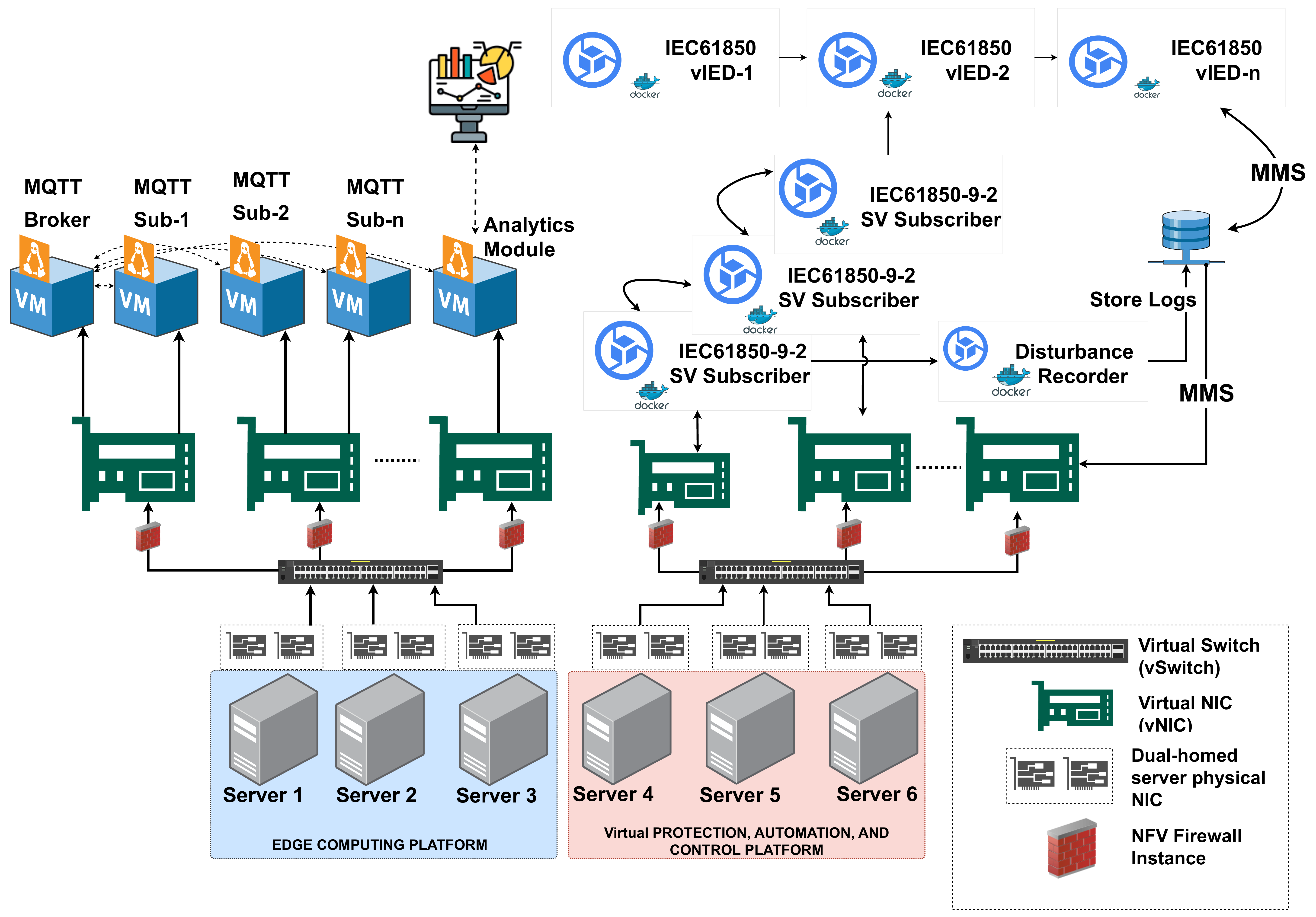}
    \caption{The server cluster design (physical layer) hosting edge computing platform (ECP) nodes and IEC61850 virtualized protection, automation, and control (vPAC) nodes.}
    \label{fig:server-cluster}
\end{figure}
Communication between WTGs and their respective OSS is established via multiplexed subsea fiber-optic cables that aggregate traffic from multiple feeder links into a high-throughput backbone, as illustrated in Figure \ref{fig:wpp}. Within each OSS, virtualized platforms host both edge computing platform (ECP) nodes and IEC 61850-compliant virtual protection, automation, and control (vPAC) nodes as illustrated in Figure \ref{fig:server-cluster}. These systems interact with turbine control units, neighboring substations, and the mainland control center as outlined in~\cite{mwangi2024dependability}.
The local area network (LAN) within each WTG and OSS control system is implemented using OpenFlow-hybrid Ethernet switches, enabling deterministic communication for both time-sensitive and best-effort traffic. In each OSS, these switches are organized into a super spine-leaf topology, connecting to dual-homed servers via active-active Linux bonding~\cite{ajibola2021}, ensuring high availability and fault tolerance. Inter-substation communication is established through line-of-sight radio links, while the OSS–mainland connection is handled via high-bandwidth export fiber cables.
To centrally manage the data plane, each OSS employs a peer-cluster of SDN controllers configured with hybrid-band control (HBC), which combines the isolation benefits of out-of-band control (OOBC) with the scalability and reduced overhead of in-band control (IBC)~\cite{carrascal2023comprehensive, sakic2020automated}. This approach provides a resilient, low-latency control architecture well suited to large-scale, latency-sensitive industrial networks.

Environmental constraints inherent to offshore deployments introduce additional design considerations. Due to harsh maritime conditions and the sensitivity of Ethernet switches to thermal stress, continuous thermal monitoring is essential. Switches are equipped with multiple sensors (e.g., inlet, outlet, CPU hotspot, power supply)~\cite{Cisco_Catalyst_9300_Datasheet,ASHRAE_TC0909_2016}, with real-time data aggregation supported by OSS management software. Thermal reliability is further improved through strategic server and equipment placement, as well as the deployment of ($n{+}1$) redundant HVAC units per OSS. 
The unique operational and environmental constraints of this clustered offshore WPP architecture prove the need for intelligent, adaptive, and resilient control mechanisms capable of maintaining service continuity under dynamic network conditions and hardware stress. Given the hierarchical yet distributed nature of the OSS-WTG topology and the use of programmable networking infrastructure, this application scenario is well suited to an externally adaptable self-healing agent. 

%%%%%%%% Self-healing framework

\section{SYSTEM MODEL} \label{model}
This section formalizes the system model and defines the problem formulation for the design and evaluation of the proposed DQN-based self-healing agent.
\medskip
\subsection{The \textit{``Observe-Orient-Decide-Act"} self-healing framework}

This paper proposes an externally adapted \textit{``Observe-Orient-Decide-Act''} self-healing framework, implemented in the Knowledge Plane shown in Figure \ref{fig:wpp}.
This \textit{``Observe-Orient-Decide-Act"}  self-healing framework performs the following functions: 
\begin{itemize}
    \item obtaining \textit{``user intents''} which are the desired outcomes or needs that guide the network design and service management from the O\&M personnel, business executives, regulators, and other key stakeholders \cite{de2021clara}, 
    \item converting these \textit{``user intents''} into \textit{``system intents''}in the ABSTRACTION Module, which is high-level network configurations and policies, using the large language models (LLMs), and
    \item employing DRL approaches that learn sequential decision-making for network resource optimization in multiple domains and application scenarios.
\end{itemize} 
The self-healing framework modules are described below:
\subsubsection{OBSERVE}
\begin{figure}
    \centering
    \includegraphics[width=\columnwidth]{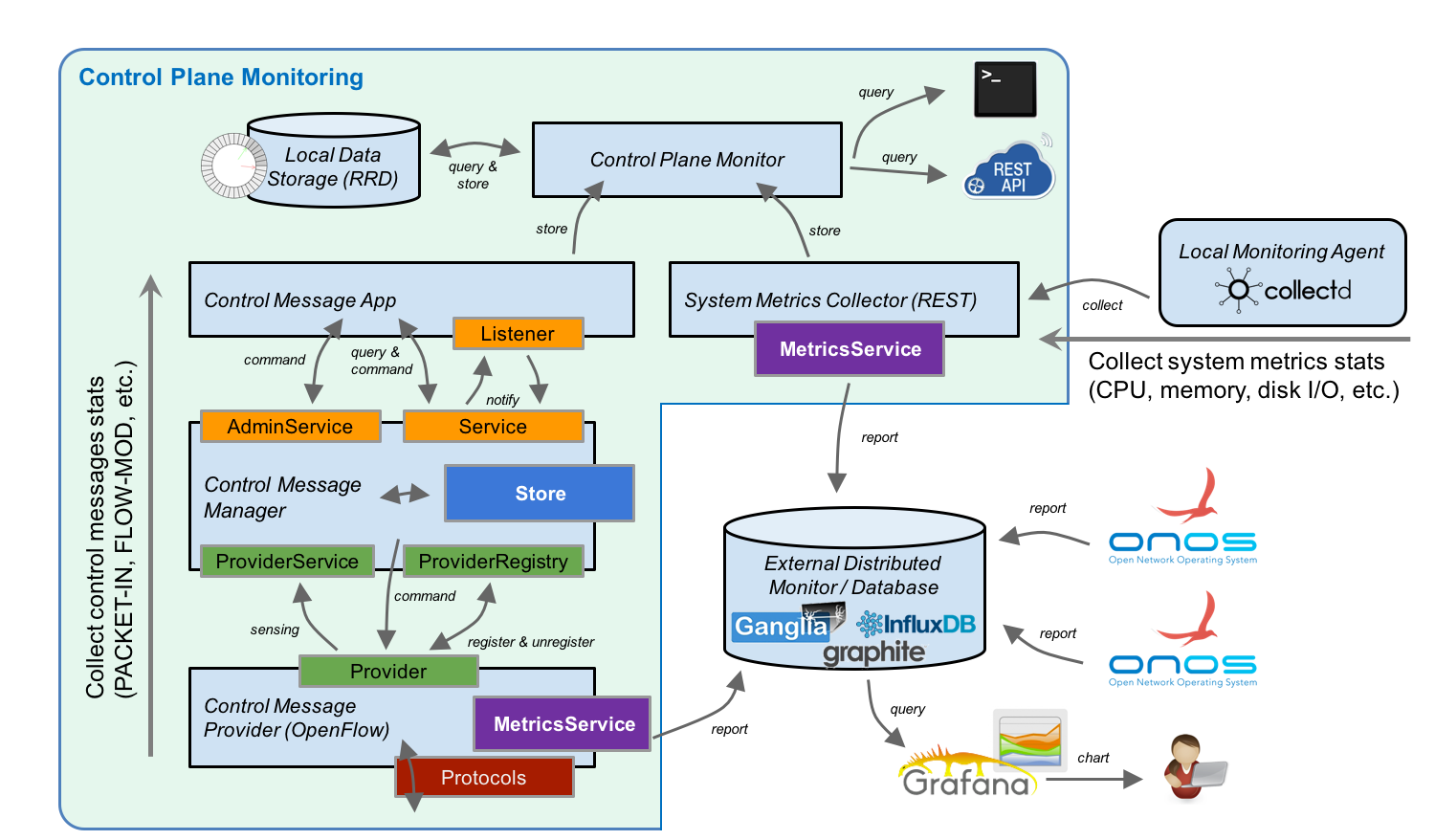}
    \caption{The ``Observe-Orient-Knowledge Base'' self-healing framework modules' schematic showing interactions between control plane, storage tools, and visualization tools \cite{ONOS_REST_API}.}
    \label{fig:ookb}
\end{figure}

The \texttt{OBSERVE} module interacts with the SDNC through the Northbound RESTful API endpoints (\textit{with OAuth 2.0 access token for secure authentication and authorization}) \cite{ONOS_REST_API} to access fine-grained metrics from the SDNC's topology manager, flow rule manager, and statistics module as illustrated in Figure \ref{fig:ookb}. In this study, the \texttt{OBSERVE} module extracts the topology and traffic matrix, yielding a non-oriented graph, as denoted in equation \ref{gve}.
\begin{equation}
    \mathcal{G}=(\mathcal{V},\mathcal{E})
    \label{gve}
\end{equation}
where $\mathcal{V}$ is a set of network nodes (OpenFlow-switches) and $\mathcal{E}$ is the set of network links (switch-to-switch physical links). The resulting topology matrix comprises $m$ nodes such that $||\mathcal{V}||=m$ and $n$ links such that $||\mathcal{E}|| =n$. This \texttt{OBSERVE} module also accesses flow rules ($\mathcal{F}$) that forward packets between network nodes (application identifier $\rightarrow$\textit{``org.onosproject.fwd''}). Additionally, the \texttt{OBSERVE} module accesses the permanent flows that facilitate the asynchronous communication for data service type classifications (application identifier $\rightarrow$\textit{``org.onosproject.core''}).
Subsequently, this module explores the operating conditions of $\mathcal{V}$ and $\mathcal{E}$ by assessing:
\begin{itemize}
    \item The traffic matrix, $TM_t$, is defined as 
    \begin{equation}
        TM_t = \begin{pmatrix}
        u_{1,t} & u_{2,t} & \dots & u_{n,t} \\
        l_{1,t} & l_{2,t} & \dots & l_{k,t}
        \end{pmatrix} \quad  n \neq k
        \label{tm}
    \end{equation}
    
    where $u_{i,t}$ is the link utilization of link $i$ at time, $t$ and $l_{j,t}$ is the aggregated latency of source-destination path $j$ at time $t$. This traffic matrix is derived passively by reading the SDNC's statistics manager or through Link Layer Discovery Protocol (LLDP) packet probing, a passive technique based on the software-defined latency monitoring framework in \cite{yu2015software}, initially piloted on large-scale data center networks.
     \item The switch temperature profile, $\tau_t$, is defined as
     \begin{equation} 
         \tau_t = (\tau_{1,t}, \tau_{2,t}, ... , \tau_{m,t})
         \label{temp}
     \end{equation}
     where $\tau_{k,t}$ is the aggregated switch temperature of switch $k$ at time $t$ for all the switches in the network topology. 
\end{itemize}

Finally, the \texttt{OBSERVE} module periodically stores the current network state, $(TM_t,\tau_t)$, as a time-stamped record in the knowledge base. 
\subsubsection{ORIENT} The \texttt{ORIENT} module reads the latest topology and traffic matrix from the \texttt{OBSERVE} module. It uses these two matrices to develop an annotated visual of the non-oriented graph, $\mathcal{G}$. 
Also, the module retrieves predefined QoS/SLA thresholds, $\mathcal{N}_{req}$, from the \texttt{ABSTRACTION} module to establish network performance constraints that align with the \textit{``system intents''}. These predefined QoS/SLA thresholds include path latency $\leq \mathcal{L}_{thr}$, link utilization $\leq \mathcal{U}_{thr}$, and device temperature $\leq \tau_{thr}$, which are checked against the current network state, $s_t$. After that, the ORIENT module triggers the \texttt{DECIDE} module when any or all of these thresholds are violated \cite{mwangi2023building}. 

\subsubsection{DECIDE} 
The \texttt{DECIDE} module executes the \textit{``threshold-triggered DQN self-healing agent''} upon receiving violation alerts from the \texttt{ORIENT} module to make real-time network management decisions that optimize performance and prevent the long-term impact of stochastic disruptions. As such, it takes corrective actions, such as rerouting traffic along optimal paths and prioritizing critical, time-sensitive traffic over best-effort traffic, when the switches are operating at nominal temperature levels.

\subsubsection{ACT} The \texttt{ACT} module reads the decisions made by the DQN self-healing agent, then creates high-priority flow rules and sends them to the SDNC's flow rule manager as a \textit{.json} file. The flow rule manager pushes these flow rules via the southbound interface to the respective switches, executing the actions recommended by the DQN self-healing agent.

\subsection{A Threshold-Triggered Deep Q-Network Self-Healing Agent (\texttt{TTDQSHA})}
The self-healing process is modeled using the classical Markov Decision Process (MDP) \cite{puterman1990markov,sutton2018reinforcement}.
This MDP is defined as a 4-element tuple $(\mathcal{S}, \mathcal{A},\mathcal{R},\mathcal{\gamma})$ where $\mathcal{S}$ is a set of the network states obtained by the \texttt{OBSERVE} module, $\mathcal{A}$ is a set of actions implemented by the \texttt{ACT} module, $\mathcal{R}$ is the immediate reward received when the system transitions from the current state $s_t$ to the next state $s_{t+1}$, and $\mathcal{\gamma}=[0,1)$ is the discount factor that enables the agent to strive for a long-term higher reward \cite{luong2019applications}.

The current network state, herein, the input state space, $s_t \in \mathcal{S}$, is defined as,
\begin{equation} \label{st}
    s_t = (TM_t, \tau_t) \in \mathbb{R}^\eta
\end{equation}

where $\mathbb{R}^\eta$ represents the $\eta$-dimensional real-valued coordinate space where $\eta=(n+k+m)$.

The action taken from the action space, $a_t \in \mathcal{A}$  comprises two actions: (i) traffic re-routing on pre-configured redundant paths and (ii) throttling low-priority traffic based on the service-type classification indicated in the flow.
\begin{equation}\label{at}
    a_t =  \begin{pmatrix}  
            p_{1,t},p_{2,t}, & \dots & , p_{i,t} \\
            f_{1,t},f_{2,t}, & \dots & , f_{j,t}
            \end{pmatrix} \quad i\neq j
\end{equation}
where $p_{w,t} $ action captures candidate or optimal paths, $w$, in source-destination pairs at time, $t$, and $f_{h,t}$ action increases the priority on the critical, time-sensitive data traffic to throttle low-priority traffic flows based on the service type classification, $h$, at time, $t$.

The reward function is designed to incentivize actions, $a_t$, that optimize network performance by minimizing end-to-end latency and overall packet loss in the current network state, $s_t$.
It is defined as,
\begin{equation} \label{rt}
    \mathcal{R}(s_t,a_t) = \mathcal{C}-(\alpha\mathbb{E}(l_{j,t}) + \beta\mathbb{E}(u_{i,t})) 
\end{equation}
where $(\alpha, \beta) \in [0,1)$ are tunable parameters representing the contributions of the expected values, $\mathbb{E}$, of the normalized network performance metrics $l_{(j,t)}$ and $u_{(i,t)}$ and $\mathcal{C}$ is the normalized link bandwidth or capacity. The reward function encourages the agent to achieve lower values of the normalized $\mathbb{E}(l_{j,t})$ and $\mathbb{E}(u_{i,t})$, while penalizing higher values.

%resulting in a well-optimized network where the performance is improved holistically. 
%See Graphical Abstract in the Images folder for better understanding of the algorithm
%See Graphical Abstract in the Images folder for better understanding of the algorithm
\RestyleAlgo{ruled}
\begin{algorithm} 
\caption{A Threshold-Triggered Deep Q-Network Self-Healing Agent}\label{alg:shm}
\textbf{Initialize:} 

\begin{enumerate}
    \item Neural Network $\mathcal{Q}$ with random weights $\theta$  %HL1 - Main neural network used to estimate the Q-values
    \item Target Neural Network $\hat{\mathcal{Q}}$ with random weights $\hat{\theta}$ %HL2 used to stabilize the learning process (updated less frequently)
    \item Experience replay memory, $\mathcal{D}$   %Buffer that stores the agent's experiences. Breaks the correlation between consecutive experiences and stabilizes learning
    %\item 
\end{enumerate}
\textbf{Step 1: }Read user-Centric (\textit{``Intents"})  defined as QoS/SLA thresholds: $\mathcal{N}_{req} = (u_{thr}, l_{thr}, \tau_{thr})$\; \Comment{\textit{Abstraction Module}} \\
\textbf{Step 2: }Read network state: $s_t$ = $(TM_t, \tau_t)$\; \Comment{\textit{Observe Module}} \\
\eIf{$ TM_t$ violates $\langle u_{thr} || l_{thr}\rangle$}{
    $\Rightarrow$ Read paths for select source-destination (src-dest) pairs\\  %\Comment{\textit{Knowledge Base}} \\
    \eIf{($\tau_{(k,t)} \geq \tau_{thr(max)}$ $||$ $\tau_{(k,t)} \leq \tau_{thr(min)}$)}{
    $\Rightarrow$ Select candidate paths for these src-dest pairs \\
    }{
    $\Rightarrow$ Load all possible src-dest pair paths\\
    \For{episode = 1,$\mathcal{M}$}{
      \[ 
    a_t = \begin{cases}
            \mathcal{A}_{(random)} & \xi \leq \epsilon, \\
            \arg\max_a Q^*(s_t,a_t,\theta) & \text{ \textit{otherwise}},
           \end{cases}
    \] 
   \textbf{Step 3: } Perform action $a_t$\;  \Comment{\textit{}\textit{Act module}} \\
   \textbf{Step 4: } Obtain reward $\mathcal{R}_{t+1}$ \;
   \textbf{Step 5: } Read the next network state $s_{t+1}$\;
   \Comment{\textit{Observe Module}} \\
   \textbf{Step 6: } Store transition ($s_t$,$a_t$,$\mathcal{R}_{t+1}$,$s_{t+1}$) in $\mathcal{D}$\;
   \textbf{Step 7: } Sample random minibatches c($s_j$,$a_j$,$r_{j+1}$,$s_{j+1}$)\;
   \For{every transition ($s_i$,$a_i$,$\mathcal{R}_{i+1}$,$s_{i+1}$) in c}{
   $y_i = \mathcal{R}_i + \gamma \max_{a' \in \mathcal{A}} \hat{Q}(s'_i, a')$
   }
   \textbf{Step 8: }Minimize the loss function: $\mathcal{L} = \frac{1}{N} \sum_{i=0}^{N-1} \left(Q(s_i, a_i) - y_i\right)^2$\;
   \textbf{Step 9: }Reset $\hat{\mathcal{Q}}=\mathcal{Q}$  every $\mathcal{M}/5$ steps\;
 }  
}
}
{
 \textbf{Step 10: }Read network state: $s_t$ = $(TM_t, \tau_t)$\; \Comment{\textit{Observe Module}}
 }   
\end{algorithm}

%%%%%%%%%%%%%%SOURCES %%%%%%%%%%%%%%%%%
% 1. https://medium.com/@devarakonda.vishnu5/simple-deep-q-learning-with-math-1afb0cfdcf0d
Algorithm \ref{alg:shm} describes the threshold-triggered Deep Q-Network self-healing model. 
%Detailed description of the algorithm.
The algorithm initializes (i) the main neural network, $\mathcal{Q}$, which predicts the action-value function, estimating the expected utility of actions for a given current state, (ii) the target neural network, $\hat{\mathcal{Q}}$ which stabilizes the learning process, and (iii) the experience replay memory, $\mathcal{D}$ which stores the agent's experiences consisting $\{s_t, a_t, \mathcal{R}(s_t,a_t), s_{t+1}\}$ as tuples in a replay memory buffer \cite{gym2021deep}. 

In step 1, the DQN self-healing agent reads the QoS/SLA requirements ($\mathcal{N}_{req}$) referred to as \textit{``system intents''} from the \texttt{ABSTRACTION} Module. 
This $\mathcal{N}_{req}$ captures the QoS/SLA thresholds the software-defined IIoT-Edge network must meet to ensure operation within acceptable limits.
In step 2, the agent reads the current network state, $s_t$.
Then, it checks if the current network state violates the stipulated QoS/SLA requirements, $\mathcal{N}_{req}$. 
If the agent detects a violation in $TM_t$, it retrieves all pre-configured paths between selected source-destination pairs from the knowledge base. It then checks whether any switches along these paths have temperatures outside the nominal operating range. The agent selects candidate paths with switches operating within the nominal temperature range and adds them to the action space.

After that, in $\mathcal{M}$ training episodes, the self-healing agent determines the best course of action by first exploring based on the epsilon-greedy policy in step 3, where $\xi \in$ \~uniform(0,1). When an action is taken, the agent obtains a reward, $\mathcal{R}_{t+1}$, based on the action taken, $a_t$, while in the state, $s_t$, in step 4 and transitions to a new state in step 5. The transition elements, $\{s_t, a_t, \mathcal{R}(s_t,a_t), s_{t+1}\}$, in this episode run are stored in the experience replay memory, $\mathcal{D}$, in step 6.
After several runs, a random mini-batch of transitions is sampled from the replay memory to update the neural network, $Q$, while preventing overfitting and stabilizing the training process, as denoted in step 7. 
The loss function expresses the difference between the current Q-value estimate $\hat{Q}(s'_i, a')$ and the target Q-value denoted in equation (\ref{eqn_qval}) as,
\begin{equation} \label{eqn_qval}
   \mathcal{L} = \frac{1}{N} \sum_{i=0}^{N-1} \left(Q(s_i, a_i) - y_i\right)^2
\end{equation}
It is minimized throughout training, making Q-value predictions more accurate and, in turn, improving the agent's decision-making.
Because the DNN is trained using an optimization algorithm, the weights, $\theta$ and $\hat{\theta}$, are updated by computing the gradient of the loss function with respect to these weights, enabling the agent to make better approximations of the true Q-values in steps 8 and 9. In step 9, the target network was updated every $\mathcal{M}/5$ steps to balance training stability and convergence speed without introducing noise into the Q-value targets.
If there is no violation, the algorithm returns to monitoring the network state in step 10.
\section{METHODOLOGY} \label{part3}
\subsection{Experimental Setup}
\begin{table}[!t]
\centering
\caption{TESTBED COMPUTE AND STORAGE RESOURCES}
\label{tab:wpp_setup_testbed}
\footnotesize
\resizebox{\columnwidth}{!}{
\begin{tabular}{|l|l|c|c|c|c|}
\hline
\textbf{Cluster / Plane} & \textbf{VM ID(s)} & \textbf{OS}$^{a}$ & \textbf{Instance}$^{b}$ & \textbf{vCPU / RAM} & \textbf{Storage}$^{c}$ \\ 
\hline
Hornsea P1 (DP) & VM-AS1-DP1 & U & E2ds-V4 & 2 @ 16GB & P20 \\
\hline
Hornsea P1 (CP) & VM-AS1-SDNC ×3 & U & E2ds-V4 & 2 @ 16GB & P20 \\
\hline
Hornsea P2 (DP) & VM-AS2-DP2 & U & E2ds-V4 & 2 @ 16GB & P20 \\
\hline
Hornsea P2 (CP) & VM-AS2-SDNC ×3 & U & E2ds-V4 & 2 @ 16GB & P20 \\
\hline
Hornsea P3 (DP) & VM-AS3-DP3 & U & E2ds-V4 & 2 @ 16GB & P20 \\
\hline
Hornsea P3 (CP) & VM-AS3-SDNC ×3 & U & E2ds-V4 & 2 @ 16GB & P20 \\
\hline
Knowledge Plane & VM-KP & W & D8lds-V5 & 8 @ 32GB & P20 \\
\hline
\end{tabular}
}

\vspace{2mm}
\begin{minipage}{\columnwidth}
\footnotesize
\textbf{KEY:}\\[1mm]
\resizebox{\columnwidth}{!}{
\begin{tabular}{@{}p{0.05\columnwidth} p{0.9\columnwidth}@{}}
$^{a}$ & \textbf{Operating System (OS):} U $\rightarrow$ Ubuntu Server 22.04 LTS (Gen2 x64); W $\rightarrow$ Microsoft Server 2022 Datacenter (Azure Edition). \\
$^{b}$ & \textbf{Instance:} Standard Azure VM instance type. \\
$^{c}$ & \textbf{Storage (P20):} Premium SSD LRS, 512 GiB, up to 2300 IOPS, 150 Mbps, server-side encryption with platform-managed keys. \\
\end{tabular}
}

\vspace{2mm}
\textbf{Additional Notes:}  
SDN control planes deployed using ONOS v2.0.0 (\textit{loon}) with the \texttt{org.onosproject.cluster-ha} feature enabled.  
Data-plane topologies emulated using Mininet 2.3.0 with OpenFlow virtual switches.
\end{minipage}
\end{table}

The proposed self-healing framework was designed, validated, and tested in a proof-of-concept testbed created on the Microsoft Azure cloud platform. The proof-of-concept testbed consisted of 15 virtual machines (VMs) with varying compute and storage specifications, as described in Table \ref{tab:wpp_setup_testbed}.
Typically, offshore WPPs consist of 100 or more WTGs linked to a single OSS. However, this experimental setup employed a reduced-scale model with 20 WTGs per WPP to reflect practical limitations in available computational and memory resources. 
Accordingly, each WPP network topology in the Mininet emulator consisted of 40 nodes, 78 links, and 60 hosts. A uniform link capacity of 1 Gbps was maintained across the virtual network.
Lastly, the \textit{``Observe–Orient–Decide–Act''} Python-scripted modules were hosted on the knowledge-plane VM to interface with the primary SDNCs across the triple-WPP cluster software-defined IIoT-Edge networks. 
Open-source tools, including \textit{collectd}, \textit{Grafana}, and \textit{InfluxDB}, were employed to collect system metrics, visualize performance trends, and store the data as time-series records, respectively.

\subsection{Traffic Generation}
In each Mininet VM, different data samples related to the offshore WPP services were generated and transmitted through the OpenFlow vSwitches in the spine-leaf network topology. Table \ref{tab:traffic-gen} presents the most common data acquisition system services for offshore WPPs \cite{mwangi2024towards} -\cite{vizarreta2019incentives}, detailing the communicating nodes categorized into \circled{i} critical and time-sensitive, \circled{ii} critical and delay-tolerant, and \circled{iii} best-effort data traffic types.

Customized publish/subscribe Python scripts were executed on randomly selected hosts tagged \textit{``Local Data Acquisition (LDAQ)''} using Eclipse\textsuperscript{\textregistered} Mosquitto (an open-source MQTT broker).
These scripts published various data types, including turbine operational, condition monitoring, meteorological, and event-driven maintenance data, to specific topics hosted by the MQTT broker running in the Mininet VM. 
This data was then subscribed to by hosts tagged \textit{``Edge Computing Platform node (ECP)''}. Simultaneously, publish/subscribe libIEC61850/SV \cite{Zillgith} were generated from hosts tagged \textit{``Merging Unit (MU)''} and subscribed to by hosts tagged \textit{``virtual Intelligent Electronic Device (vIED)''}. 
% \usepackage{multirow}
% \usepackage{graphicx}
\begin{table}
\caption{OFFSHORE WPP DATA SERVICES AND REQUIREMENTS \cite{vizarreta2019incentives,mwangi2024towards}}
\label{tab:traffic-gen}
\resizebox{\columnwidth}{!}{%
\begin{tabular}{|l|l|l|l|}
\hline
\textbf{WPP Data Service}                                                                                                        & \textbf{Nodes}         & \textbf{Sampling Rate}                                                      & \textbf{Type}                                    \\ \hline
\begin{tabular}[c]{@{}l@{}}Turbine Operational Data \\ (e.g., power output, rotor speed,\\ wind speed and direction)\end{tabular} & WTG $\rightarrow$ ECP  & 10 Hz - 1 kHz                                                                          & \multirow{3}{*}{\begin{tabular}[c]{@{}l@{}}- Critical\\ - Time sensitive\end{tabular}} \\ \cline{1-3}
Control traffic                                                                                                                  & vIED $\rightarrow$ WTG & 4800 sps                                                                             &                                                                                        \\ \cline{1-3}
Protection traffic                                                                                                               & WTG $\rightarrow$ vIED & 9600 sps                                                                 &                                                                                        \\ \hline
\begin{tabular}[c]{@{}l@{}}Condition monitoring data \\ (e.g., vibration)\end{tabular}                                            & WTG $\rightarrow$ ECP  & 1 Hz - 1 kHz                                                                          & \begin{tabular}[c]{@{}l@{}}- Critical\\ - Delay tolerant\end{tabular}                  \\ \hline
\begin{tabular}[c]{@{}l@{}}Meteorological Data\\ (e.g., temperature, humidity, \\ atmospheric pressure)\end{tabular}              & WTG $\rightarrow$ ECP  &  0.1Hz - 1Hz                                                                  & \multirow{2}{*}{- Best effort}                                                         \\ \cline{1-3}
\begin{tabular}[c]{@{}l@{}}Event-driven Maintenance Data \\ (e.g., alerts, logs, and alarms)\end{tabular}                         & WTG $\rightarrow$ ECP  & Event-driven                                                     &                                                                                        \\ \hline
\end{tabular}%
}
\end{table}

These data samples were published at varying rates as denoted in Table \ref{tab:traffic-gen}. At the same time, vPAC node subscribers polled for updates and published libIEC61850/GOOSE \cite{Zillgith}. The ECP node subscribers received notifications of new data samples from MQTT broker topics using an event-driven mechanism.
Flash events based on WPP operation and maintenance activities were generated by running \textit{``iperf3''} commands that sent large Transmission Control Protocol (TCP) and User Datagram Protocol (UDP) packets through the switch network.

At the \textit{knowledge plane}, the \texttt{OBSERVE} module continuously collects real-time network metrics. Using ONOS’s native telemetry APIs, it retrieves the traffic matrix $TM_t$ (in bits per second) every $0.1$~ms. The \texttt{ACT} module then generates the required flow rules in \texttt{.json} format and sends them to the SDNC’s Flow Rule Manager, which asynchronously installs the updates on the switch data plane using \texttt{OpenFlow\_MOD} messages.
The proposed \texttt{TTDQSHA} framework demonstrates substantially improved responsiveness. Its \texttt{ORIENT} module detects SLA violations within $0.1$--$0.3$~ms, and the subsequent flow insertion performed by the \texttt{ACT} module completes within $1$--$7.8$~ms, as observed via Wireshark\textsuperscript{\textregistered} captures.

\subsection{Temperature Module}
Lastly, a temperature module was modeled as a lumped-capacity, first-order thermal model \cite{erden2014hybrid,salih2014determination,demetriou2014development} to capture the inherent thermal inertia and lag observed in physical data-plane network hardware, since Mininet does not expose hardware-level temperature measurements. Further, the module benchmarks the behaviour of industrial Ethernet switching hardware \cite{Cisco_Catalyst_9300_Datasheet}. 
In accordance with ASHRAE TC9.9 (Class A1/A2) guidelines \cite{ASHRAE_TC0909_2016}, the recommended inlet temperature for network equipment ranges from approximately $18–27^\circ$C. 
This recommended envelope provides the baseline for modeling the ambient inlet temperature supplied to each switch in an offshore substation environment. 
The referenced standards and data sheets specify a nominal operating ambient temperature range of approximately $-5^\circ$C to $+45^\circ$C at sea level, with altitude-dependent derating reducing the upper limit to around $+35^\circ$C at 15000~ft \cite{Cisco_Catalyst_9300_Datasheet}. 
These thresholds define the allowable limits for ambient inlet temperature delivered to the switch.
Two coupled thermal states are modeled: \circled{1} ambient (inlet) temperature for switch, $k$, denoted  $\tau_{k}^{(ambient)}(t)$, representing the airflow temperature delivered to the switch in the data center environment, as denoted in equation \ref{eq:ambient_temp},
\begin{equation}
\begin{aligned}
\frac{d \tau^{\mathrm{(ambient)}}_{k}(t)}{dt}
&=
\frac{1}{\lambda_{\mathrm{ambient}}}
\left(
\tau^{\mathrm{env}}(t) - \tau^{\mathrm{ambient}}_{k}(t)
\right) \\
&\quad
+ \kappa_{\mathrm{rack}}\,P_{\mathrm{rack}}(t)
- \kappa_{\mathrm{cool}}\,C_{\mathrm{hvac}}(t)
\end{aligned}
\label{eq:ambient_temp}
\end{equation}
where $\tau^{\mathrm{ambient}}_{k}(t)$ denotes the ambient (inlet) air temperature supplied to switch $k$ at time $t$, $\tau^{\mathrm{env}}(t)$ is the external offshore environmental temperature, $\lambda_{\mathrm{ambient}}$ represents the ambient thermal time constant associated with rack and HVAC thermal inertia, $P_{\mathrm{rack}}(t)$ is the normalized rack-level IT load that contributes to heat generation, $C_{\mathrm{hvac}}(t)$ denotes the effective HVAC cooling level (with 0 indicating degraded cooling and 1 indicating full cooling capacity), while $\kappa_{\mathrm{rack}}$ and $\kappa_{\mathrm{cool}}$ are thermal gain parameters describing the influence of rack load and HVAC cooling on the ambient inlet temperature, respectively.
\circled{2} the internal temperature profile, $\tau_{k}^{(internal)}(t)$ resulting from two contributing factors: \circled{1} idle baseline temperature rise resulting from chassis electrons, fans, and the base forwarding ASIC, even with minimal network traffic, and \circled{2} load-dependent heating driven by packet processing, forwarding ASIC activity, and CPU utilization as network traffic increases, as denoted in equation \ref{eq:internal_temp},
\begin{equation}
\begin{aligned}
\frac{d \tau_{k}^{(internal)}(t)}{dt}
&=
\frac{1}{\lambda_{\mathrm{sw}}}
\left(
\tau^{\mathrm{ambient}}_{k}(t) + \psi_{\mathrm{idle}} - \mathcal{U}_{k}(t)
\right) \\
&\quad
+ \phi_{\mathrm{sw}}\,\mathcal{U}_{k}(t)
\end{aligned}
\label{eq:internal_temp}
\end{equation}
\noindent\textit{where} $\tau_{k}^{(internal)}(t)$ denotes the internal temperature of switch $k$ at time $t$, $\tau^{\mathrm{ambient}}_{k}(t)$ is the ambient (inlet) temperature supplied to the switch, $\lambda_{\mathrm{sw}}$ represents the internal thermal time constant capturing the thermal inertia of the chassis, heatsinks, and airflow, $\psi_{\mathrm{idle}}$ denotes the idle temperature rise above the inlet air associated with baseline power consumption, $\mathcal{U}_{k}(t)$ is the normalized switch utilization derived from network load polled from the knowledge base, and $\phi_{\mathrm{sw}}$ is the thermal gain parameter that maps utilization to additional load-dependent heat generation.

\subsection{Implementing the threshold-triggered Deep Q-Network (DQN) self-healing agent in the knowledge plane}

\begin{table}
\centering
\caption{HYPERPARAMETERS USED FOR TRAINING THE THRESHOLD-TRIGGERED DQN SELF-HEALING AGENT}
\label{tab:hyperparameters}
\renewcommand{\arraystretch}{1.15}
\begin{tabular}{p{0.42\columnwidth} p{0.45\columnwidth}}
\hline
\textbf{Parameter} & \textbf{Description} \\
\hline
Framework & TensorFlow~2.16.2  \\
Network Models & \textit{Q-Network} and \textit{Q-Target} \\
State size & $|\mathcal{S}|$ \\
Action space size & $|\mathcal{A}|$\\
Hidden Layer Neurons & 2@(24) (\textit{ReLU} activation) \\
Output Layer & Linear activation (Q values)\\
Replay Buffer Type & \textit{deque} \\
Replay Buffer Capacity & 2000 transitions \\
Mini-Batch Size & 32 \\
Discount Factor ($\gamma$) & 0.995 \\
Learning Rate (Adam) & 0.001 \\
Random Seeds & \{23, 37, 49, 71\}; default seed: 42 \\
Exploration Rate ($\epsilon$) & 1.0 $\rightarrow$ 0.01 (decay 0.995) \\
Reward Shaping Weights & $\alpha=0.657$, $\beta=0.345$ (\textit{tuned}) \\
Target Update Frequency & Every $300$ training steps  \\
Training Episodes ($\mathcal{M}$) & 1500 \\

\hline
\end{tabular}
\end{table}
The proposed \texttt{TTDQSHA} was implemented in the Anaconda 2.6.3 environment, and key hyperparameters and architectural details used for training and testing are summarized in Table~\ref{tab:hyperparameters}. 
%Reviewer 5: - It is necessary to explain the decisions related to the number of neurons used in the NNs and the use of the ReLU as an activation function.

Two sequential deep neural networks were employed: the \textit{Q-Network} and the periodically updated \textit{Q-Target} model. 
These parameters were tuned empirically to ensure stable convergence, efficient exploration, and sustained performance across varying IIoT-Edge network states.
\begin{figure}
    \centering
    \includegraphics[width=0.8\columnwidth]{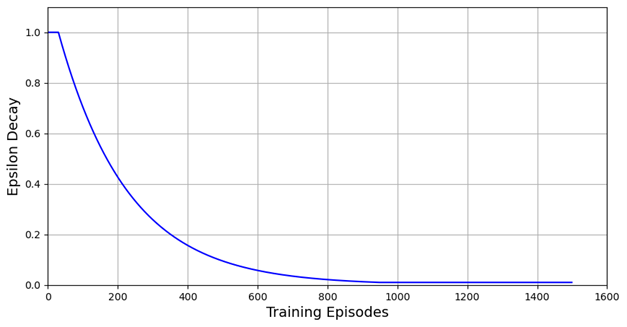}
    \caption{Learning rate and epsilon decay over the training episodes.}
    \label{fig:epsilon}
\end{figure}
%Reviewer 5: - It is necessary to explain how the decisions related to the size of the memory buffer and the de-queuing used were made.
At the beginning of the training process, the DQN self-healing agent operated with a high exploration rate ($\epsilon=1.0$), which facilitated broad exploration of the action space. 
During this early phase, performance was below optimal as the agent explored various network states and learned the consequences of its decisions. 
Over successive episodes, $\epsilon$ decayed by a factor of 0.995 until it reached a steady-state value of 0.01, allowing the agent to gradually shift from exploration to exploitation as it gained more experience and confidence in its learned policy.

This decay process, illustrated in Figure~\ref{fig:epsilon}, captures the agent’s transition from random exploration to strategic decision-making as defined in Algorithm~\ref{alg:shm}. 
The learning rate for the Adam optimizer was set to 0.001 to control the step size during gradient-descent updates, while the \textit{Q-Target} network was periodically synchronized with the \textit{Q-Network} to provide a stable reference for learning.
The replay memory was implemented using a \textit{deque}, which stores and replaces experiences in a First-In–First-Out manner to enable efficient sampling. A buffer size of 2000 transitions was chosen empirically because it provided a good balance among learning stability, sample diversity, and computational efficiency, consistent with common DQN practice.
As the exploration rate decreased, the agent increasingly relied on the optimal strategies it had acquired, leading to progressively improved decision-making and network adaptation. 

\begin{figure}
    \centering
    \includegraphics[width=\columnwidth]{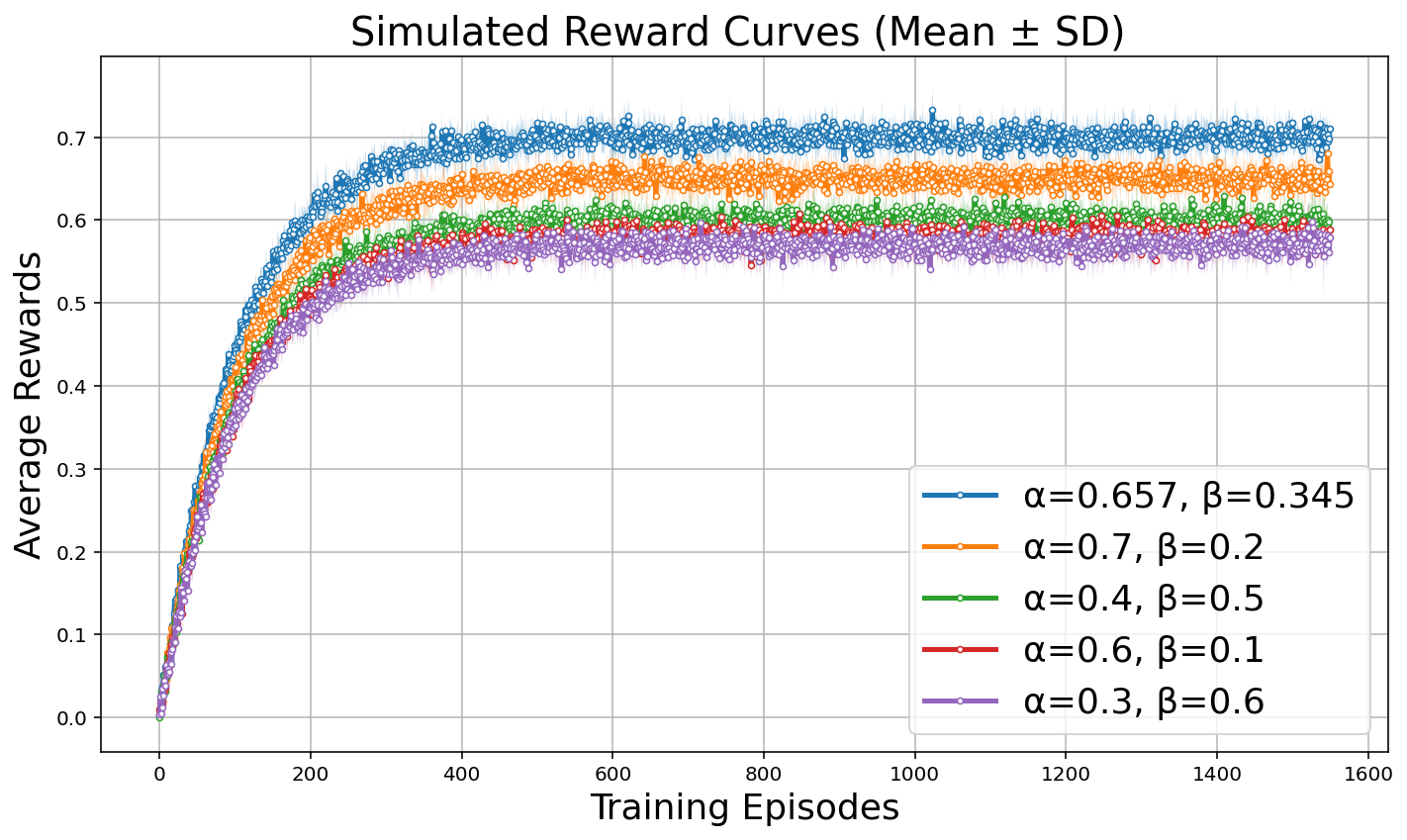}
    \caption{Reward curve showing accumulated rewards over the training episodes.}
    \label{fig:rewardcurve}
\end{figure}

%The random seeds were defined for rigor and self-aware runs. The choice of seeds was based on the Hitchhiker's Guide to the Galaxy, zero bias, and mathematical Easter eggs (pi, e). 
To determine appropriate values for the reward-shaping coefficients $\alpha$ and $\beta$, several combinations were empirically tested to evaluate their effects on the convergence and stability of the proposed TTDQSHA during training. As shown in Figure~\ref{fig:rewardcurve}, five candidate pairs were compared over 1500 training episodes, with each configuration representing a different trade-off between latency minimization and packet-loss reduction. The combination $\alpha = 0.657$ and $\beta = 0.345$ yielded the highest cumulative reward and the smoothest convergence (95\% confidence interval across random seeds), indicating a balanced trade-off between end-to-end latency and overall packet reliability. Based on this empirical hyperparameter tuning, this combination was adopted for the final implementation in the TTDQSHA model. 

Initially, rewards were relatively low when $\epsilon=1.0$, since actions were largely random. 
However, as the agent’s policy matured, the accumulated rewards increased significantly, demonstrating its enhanced ability to manage traffic rerouting, prioritization, and resource allocation across diverse network states, capabilities essential for sustaining latency-sensitive WPP data services.

\begin{figure}
    \centering
    \subfigure[Compute resource consumption during training.]
    {
        \includegraphics[width=0.75\columnwidth]{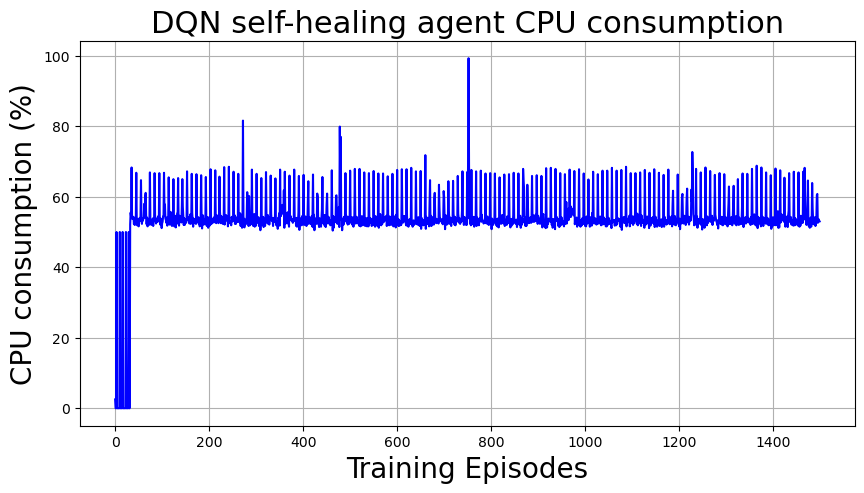}
        \label{fig:cpu}
    }
    \subfigure[Memory resource consumption during training.]
    {
        \includegraphics[width=0.75\columnwidth]{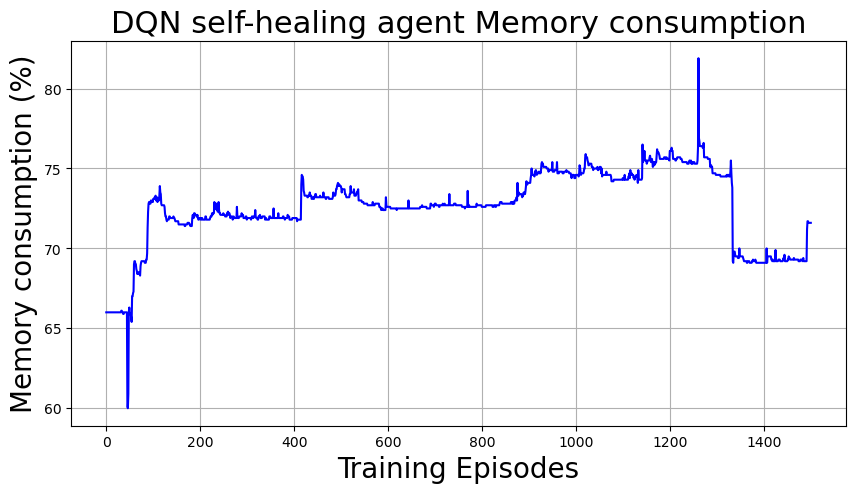}
        \label{fig:mem}
    }
    \caption{Testbed resource usage: compute and memory consumption over the training period.}
    \label{fig:cpu-mem}
\end{figure}

In addition to learning efficiency, computational performance was also evaluated. 
The CPU and memory utilization during training, depicted in Figure~\ref{fig:cpu-mem}, confirm that the DQN algorithm operated efficiently within the testbed comprising 15 virtual machines modeling a reduced-scale WPP network. 
Resource usage remained within acceptable limits throughout the training period, demonstrating both scalability and suitability for deployment in resource-constrained industrial environments. 
Based on the observed consumption of the \texttt{DECIDE} module’s VM, the minimum hardware requirements for deploying the self-healing agent are estimated at 12~GB of RAM and 2~vCPUs, given that average resource utilization ranged between $65\%$ and $80\%$.

\begin{table*}[h!]
\caption{SCENARIO MATRIX FOR EVALUATING NETWORK BEHAVIOUR AND SWITCH THERMAL DYNAMICS UNDER VARYING TRAFFIC LOAD AND ENVIRONMENTAL CONDITIONS.}
\label{tab:thermal_testcases}
\centering
\begin{tikzpicture}
\shade[left color=green!50, right color=red!50, shading=axis] (0,0) rectangle (5,0.5);
\node[align=center] at (0,-0.2) {Healthy Cases};
\node[align=center] at (5,-0.2) {Extreme Cases};
\vspace{1.5em}
\end{tikzpicture}
\vspace{0.35cm}
\resizebox{0.9\textwidth}{!}{%
\begin{tabular}{|l|l|l|l|l|l|l|l|l|l|l|}
\hline
\textbf{Test Case} &
  \textbf{$\tau_{\mathrm{ambient}}$} &
  \textbf{$\tau_{\mathrm{internal}}$} &
  \textbf{$\mathcal{U}_{(i,t)}$} &
  \textbf{$\mathcal{L}_{(j,t)}$} &
  \textbf{$\lambda_{\mathrm{ambient}}$} &
  \textbf{$\lambda_{\mathrm{sw}}$} &
  \textbf{$\kappa_{\mathrm{rack}}$} &
  \textbf{$\kappa_{\mathrm{cool}}$} &
  \textbf{$\psi_{\mathrm{idle}}$} &
  \textbf{$\phi_{\mathrm{sw}}$} \\ \hline

TC1 &
  \cellcolor{green!30}{[}18,27{]} &
  \cellcolor{green!30}{[}20,40{]} &
  \cellcolor{green!30}$\ll$80\% &
  \cellcolor{green!30}$\ll$3ms &
  \cellcolor{green!30}300.0 &
  \cellcolor{green!30}200.0 &
  \cellcolor{green!30}0.80 &
  \cellcolor{green!30}1.20 &
  \cellcolor{green!30}5.0 &
  \cellcolor{green!30}12.0 \\ \hline

TC2 &
  \cellcolor{green!30}{[}18,27{]} &
  \cellcolor{yellow!30}{[}20,45{]} &
  \cellcolor{yellow!30}$\sim$80\% &
  \cellcolor{yellow!30}$\sim$3ms &
  \cellcolor{green!30}310.0 &
  \cellcolor{green!30}210.0 &
  \cellcolor{yellow!20}0.82 &
  \cellcolor{yellow!20}1.15 &
  \cellcolor{green!30}5.1 &
  \cellcolor{yellow!20}12.5 \\ \hline

TC3 &
  \cellcolor{red!30}$\ll$18 &
  \cellcolor{yellow!30}{[}25,35{]} &
  \cellcolor{green!30}$\ll$80\% &
  \cellcolor{green!30}$\ll$3ms &
  \cellcolor{yellow!30}330.0 &
  \cellcolor{yellow!20}230.0 &
  \cellcolor{green!30}0.80 &
  \cellcolor{yellow!20}1.10 &
  \cellcolor{yellow!20}5.3 &
  \cellcolor{green!30}12.0 \\ \hline

TC4 &
  \cellcolor{red!30}$\gg$27 &
  \cellcolor{yellow!30}{[}20,50{]} &
  \cellcolor{green!30}$\ll$80\% &
  \cellcolor{green!30}$\ll$3ms &
  \cellcolor{yellow!30}340.0 &
  \cellcolor{yellow!30}240.0 &
  \cellcolor{yellow!30}0.85 &
  \cellcolor{yellow!30}1.05 &
  \cellcolor{yellow!30}5.4 &
  \cellcolor{yellow!30}12.5 \\ \hline

TC5 &
  \cellcolor{red!30}$\ll$18 &
  \cellcolor{orange!30}{[}25,40{]} &
  \cellcolor{orange!30}$\ge$80\% &
  \cellcolor{yellow!30}$\sim$3ms &
  \cellcolor{orange!30}360.0 &
  \cellcolor{orange!30}260.0 &
  \cellcolor{yellow!30}0.90 &
  \cellcolor{orange!20}0.95 &
  \cellcolor{orange!30}5.8 &
  \cellcolor{orange!30}13.0 \\ \hline

TC6 &
  \cellcolor{red!30}$\gg$27 &
  \cellcolor{orange!30}{[}30,55{]} &
  \cellcolor{orange!30}$\ge$80\% &
  \cellcolor{yellow!30}$\sim$3ms &
  \cellcolor{orange!30}380.0 &
  \cellcolor{orange!30}280.0 &
  \cellcolor{orange!30}0.95 &
  \cellcolor{orange!30}0.90 &
  \cellcolor{orange!30}6.0 &
  \cellcolor{orange!30}13.5 \\ \hline

TC7 &
  \cellcolor{green!30}{[}18,27{]} &
  \cellcolor{red!30}{[}20,45{]} &
  \cellcolor{red!30}$\gg$80\% &
  \cellcolor{orange!30}$\ge$3ms &
  \cellcolor{red!20}420.0 &
  \cellcolor{orange!30}300.0 &
  \cellcolor{orange!30}1.00 &
  \cellcolor{red!20}0.80 &
  \cellcolor{orange!30}6.5 &
  \cellcolor{red!20}14.0 \\ \hline

TC8 &
  \cellcolor{red!30}$\ll$18 &
  \cellcolor{red!30}{[}30,55{]} &
  \cellcolor{red!30}$\gg$80\% &
  \cellcolor{red!30}$\gg$3ms &
  \cellcolor{red!30}450.0 &
  \cellcolor{red!20}330.0 &
  \cellcolor{red!30}1.10 &
  \cellcolor{red!30}0.70 &
  \cellcolor{red!20}7.0 &
  \cellcolor{red!20}14.5 \\ \hline

TC9 &
  \cellcolor{red!30}$\gg$27 &
  \cellcolor{red!30}{[}30,55{]} &
  \cellcolor{red!30}$\gg$90\% &
  \cellcolor{red!30}$\gg$5ms &
  \cellcolor{red!40}500.0 &
  \cellcolor{red!40}380.0 &
  \cellcolor{red!40}1.20 &
  \cellcolor{red!40}0.60 &
  \cellcolor{red!40}8.0 &
  \cellcolor{red!40}15.0 \\ \hline

\end{tabular}
}
\vspace{2mm}
\begin{minipage}{0.9\textwidth}
\footnotesize
\textbf{Varied Thermal Constants (and Causes):}\\[2mm]
\begin{tabular}{@{}|p{0.1\textwidth} |p{0.84\textwidth}|@{}}
\hline
     $\lambda_{\mathrm{ambient}}$ & Rack and HVAC inertia affected by partial clogging, salt spray, blocked vents, and HVAC degradation \\ 
     $\lambda_{\mathrm{sw}}$      & Increased internal thermal inertia due to clogged fans, fouling, corrosion, bearing wear, and reduced fan speed \\ 
     $\kappa_{\mathrm{rack}}$     & Increased thermal gain from rack load due to degraded heat dissipation, dust accumulation, and reduced airflow efficiency \\ 
     $\kappa_{\mathrm{cool}}$     & Reduced HVAC cooling effectiveness caused by filter clogging, duct corrosion, fouled intake meshes, or fan degradation \\ 
     $\psi_{\mathrm{idle}}$       & Elevated idle baseline heat caused by hardware aging, VRM inefficiency, dust buildup, and degraded heatsink performance \\ 
     $\phi_{\mathrm{sw}}$         & Increased load-dependent heat generation due to ASIC aging, higher power per packet, increased energy per bit, and thermal inefficiency \\ \hline
\end{tabular}
\end{minipage}

\end{table*}

%Traffic Matrix
%https://core.ac.uk/download/pdf/231922994.pdf
%https://infohub.delltechnologies.com/en-us/l/tech-book-dell-integrated-system-for-microsoft-azure-stack-hci-1/overview-5441/
%https://chrisjhart.com/Install-iperf3-on-Ubuntu-22.04/
%https://git-scm.com/book/en/v2/Getting-Started-First-Time-Git-Setup
%Figure \ref{fig:poc} illustrates the experimental setup on the Microsoft Azure platform. \textit{Indicate which schemes such as the Dijkstra algorithm will be compared to my novel approach.}}

To evaluate the performance of \texttt{TTDQSHA} under stochastic disruptions \circled{A} and \circled{B}, we compared it against three state-of-the-art benchmark self-healing agents. For all these benchmarks, this study used the telemetry rates and parameters defined in their original designs to ensure methodological fidelity.
\texttt{(1) Baseline (Dijkstra+ECMP):} This conventional super spine–leaf configuration uses deterministic Dijkstra routing and ECMP forwarding \cite{rhamdani2018equal}. It is implemented in ONOS using \texttt{org.onosproject.fwd} (reactive forwarding) and \texttt{org.onosproject.pathpainter}. Network-state information is obtained via \texttt{PACKET\_IN} events and ONOS’s default $\approx$2\,s polling. The baseline has no thermal observability, hence was not evaluated for stochastic disruption \circled{B}. Also, its anomaly detection time typically ranged from 50\,ms to several seconds \cite{wang2023data}.
\texttt{(2) ANFIS:} The adaptive neuro-fuzzy inference system self-healing agent adapted from \cite{gurumekala2022toward} was evaluated using the original input features, membership functions, and tuning setup. ANFIS used the original 100 ms telemetry interval, five membership functions per input, and a rule base of 25 fuzzy rules. \texttt{(3) DTPRO:} The Deep Q-Network and traffic-prediction routing method adapted from \cite{bouzidi2021deep} was implemented using its original configuration. DTPRO collects link throughput, per-flow size, and active-probe latency every 5\,s via ONOS’s \emph{Statistics} and \emph{Latency Measurement} modules. Its DQN agent follows the published two-layer dense architecture (state size = 48), with an action space of 120 link-weight configurations, a replay memory of 500 entries, a mini-batch size of 32, a learning rate of 0.01, a discount factor of 0.95, and target-network synchronization every 300 steps. The agent updates its routing strategy every $T_{\text{DQN}} = 1$\,h, and routing actions are produced via the original mixed-integer linear-programming formulation. As in the original design, DTPRO addresses stochastic disruption \circled{A} only.

%To ensure a transparent and reproducible comparison, Table~\ref{tab:benchmark-summary} summarizes the polling cadences, observability, and configuration parameters used for each benchmark controller. All parameter settings for ANFIS and DTPRO are preserved exactly as reported in their original publications, while the baseline ECMP configuration follows ONOS’s default reactive-forwarding design. This table makes explicit the information available to each controller and clarifies the conditions under which the comparative evaluation is performed.
%\input{Tables/DQN_params}

\section{EMPIRICAL ANALYSIS} \label{part4}

This section presents the results of the proposed threshold-triggered DQN self-healing agent's training process, the network performance evaluation, and a discussion highlighting industrial application recommendations and insights into future work.

\subsection{Network Performance Assessment and Scenario-Based Sensitivity Analysis}

\begin{figure}
    \centering
    \includegraphics[width=\columnwidth]{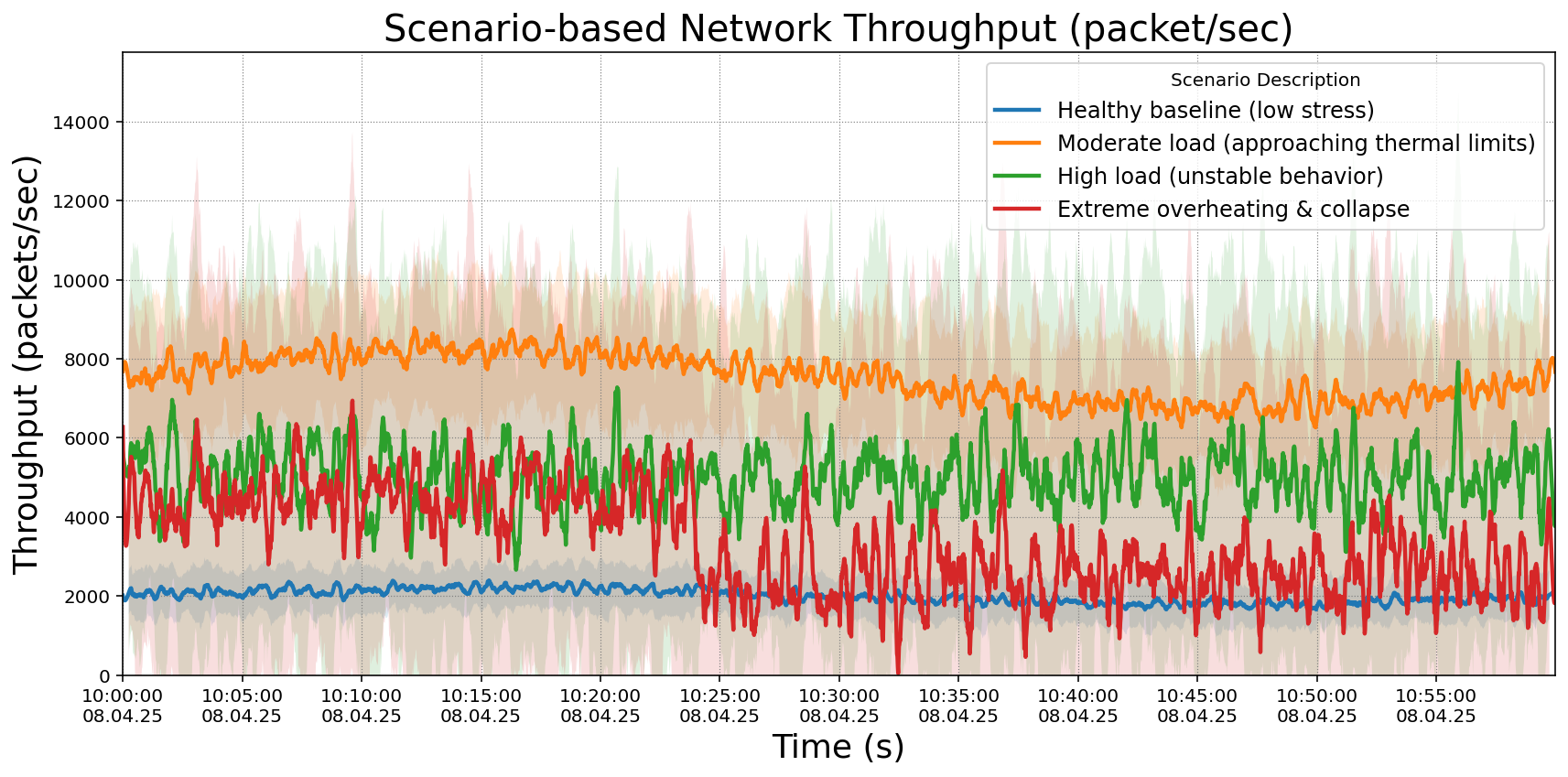}
    \caption{Scenario-based network throughput showing throughput spike during flash events (packets/sec) on a resource-constrained network.}
    \label{fig:febt}
\end{figure}

For the baseline approach, benign traffic flows were simulated using \texttt{iperf3} in Mininet topologies for Hornsea P1, P2, and P3. The scenario-based \texttt{iperf3} throughput traces (see Figure~\ref{fig:febt}) illustrate how network behavior evolves with increasing thermal and load stress across representative test cases, highlighting clear distinctions among healthy operation, pre-thermal-limit conditions, unstable high-load behavior, and extreme overheating that leads to collapse. Across these scenarios, \texttt{iperf3} reported datagram loss ranging from $3.1\%$ to $5.3\%$, with end-to-end latency rising from  $\approx 1.6$~ms under healthy conditions to peaks of up to $10$~ms during severe thermal and congestion events.
From the \texttt{ONOS CP-Manager} network monitoring dashboard, it was observed that the switches maintained a consistently high volume of \texttt{REPLY} messages ($\approx 9900$), indicating regular, stable interaction with the controller. The \texttt{FLOW-MOD} counts remained moderate (350--400), suggesting limited changes to flow entries after initial configuration. \texttt{INBOUND} and \texttt{OUTBOUND} traffic showed periodic peaks, reflecting synchronized control actions or status polling. The low \texttt{REQUEST} counts ($\approx 78$) further imply that most flows were managed proactively rather than installed reactively. Importantly, communication between the northbound interface between the ACT and the flow rule manager of the control plane exhibited latencies in the range of approximately 0.0135 to 0.0562~$\mu$s. In contrast, the southbound interface control messages (e.g., \texttt{OpenFlow\_MOD}) between the SDNC and switches occur within 0.05 to 0.2~$\mu$s. These ultra-low latencies ensure near-real-time transfer of control messages, enabling rapid flow rule installation and efficient network management. Table~\ref{tab:tc_results} summarizes these observations quantitatively, showing how latency, packet loss, throughput, and reaction time degrade progressively across the nine thermal–load test cases.

\begin{figure}
    \centering
    \subfigure[Mean end-to-end latency (ms) across TC1–TC9.]
    {
        \includegraphics[width=\columnwidth]{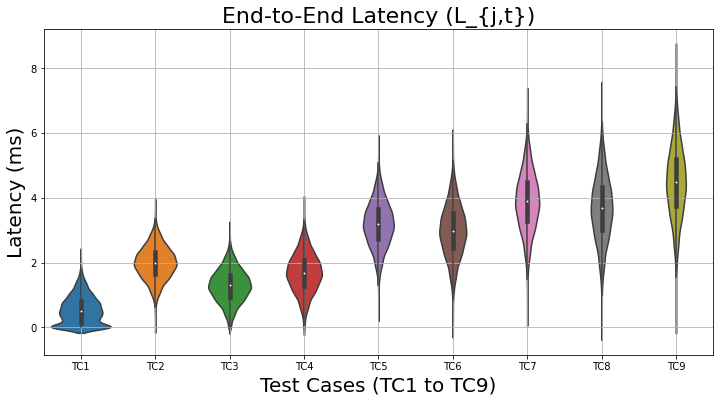}
        \label{fig:e2elat}
    }
    \subfigure[Mean packet loss ratio across TC1–TC9.]
    {
        \includegraphics[width=\columnwidth]{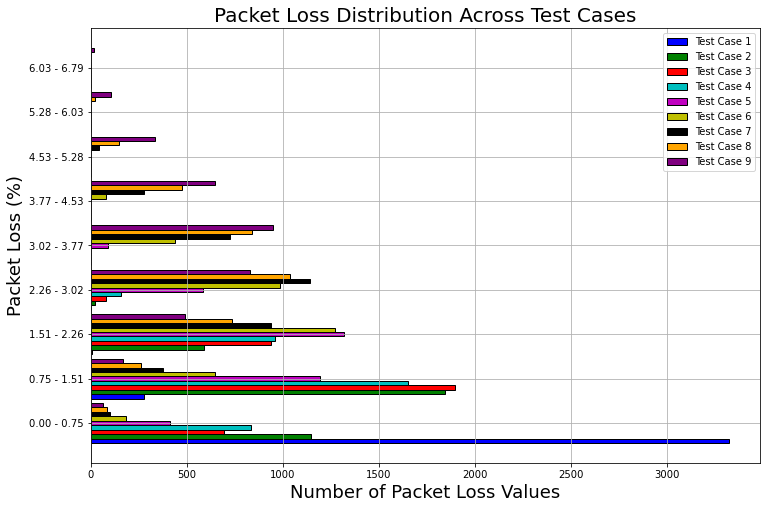}
        \label{fig:ploss}
    }
    {\caption{Mean network performance across TC1–TC9 over five independent runs.}
}
    \label{fig:net_perf}
\end{figure}

\begin{table}
\centering
\caption{AGGREGATED PERFORMANCE METRICS ACROSS TEST CASES (MEAN ± 95\% CI ACROSS 5 SEEDS)}
\label{tab:tc_results}
\resizebox{\columnwidth}{!}{
\begin{tabular}{|c|c|c|c|c|}
\hline
\textbf{Test Case} & 
\textbf{Latency (ms)} & 
\textbf{Packet Loss (\%)} & 
\textbf{Throughput (Mbps)} &
\textbf{Reaction Time (s)} \\ \hline

TC1 & $0.535 \pm 0.024$ & $0.209 \pm 0.011$ & $62.4 \pm 0.37568$ & $0.35 \pm 0.02$ \\ \hline
TC2 & $1.989 \pm 0.069$ & $1.011 \pm 0.029$ & $56.0 \pm 0.92689$ & $0.78 \pm 0.05$ \\ \hline
TC3 & $1.290 \pm 0.034$ & $1.207 \pm 0.054$ & $49.6 \pm 0.47811$ & $1.12 \pm 0.07$ \\ \hline
TC4 & $1.687 \pm 0.065$ & $1.213 \pm 0.051$ & $44.8 \pm 0.14151$ & $1.44 \pm 0.09$ \\ \hline
TC5 & $3.184 \pm 0.154$ & $1.611 \pm 0.044$ & $41.6 \pm 0.91712$ & $2.06 \pm 0.11$ \\ \hline
TC6 & $2.982 \pm 0.153$ & $2.110 \pm 0.039$ & $38.4 \pm 0.69273$ & $2.48 \pm 0.15$ \\ \hline
TC7 & $3.880 \pm 0.141$ & $2.510 \pm 0.090$ & $35.2 \pm 0.46833$ & $3.21 \pm 0.18$ \\ \hline
TC8 & $3.678 \pm 0.042$ & $2.811 \pm 0.139$ & $32.8 \pm 0.30004$ & $3.89 \pm 0.20$ \\ \hline
TC9 & $4.476 \pm 0.230$ & $3.212 \pm 0.116$ & $30.4 \pm 0.13174$ & $4.63 \pm 0.25$ \\ \hline

\end{tabular}
}
\end{table}

Across the nine test cases (TC1–TC9), a consistent pattern emerged: the DQN self-healing framework outperformed the baseline in end-to-end latency, packet loss, and thermal management. The baseline exhibited significant spikes in latency, particularly under stress conditions in TC2, TC6, and TC9, where latency exceeded 9 ms. In contrast, as illustrated in Figure \ref{fig:e2elat}, the DQN-controlled routing maintained latency below 3~ms even under high utilization, closely adhering to the defined SLA threshold $\mathcal{L}_{thr} < 3$~ms.
Packet loss followed a similar trend as illustrated in Figure \ref{fig:ploss}. Under flash traffic conditions (such as TC1 and TC5), baseline routing showed packet loss ranging from 3.1\% to 5.3\%. 
The proposed \texttt{TTDQSHA} mitigated this by proactively reassigning flows and prioritizing delay-sensitive traffic using application-layer identifiers (ETHTypes), reducing packet loss below 1.2\% in all cases. This improvement is critical for WPPs, where control messages such as IEC61850 GOOSE and SV must maintain high reliability.

\begin{figure}
    \centering
    \includegraphics[width=\columnwidth]{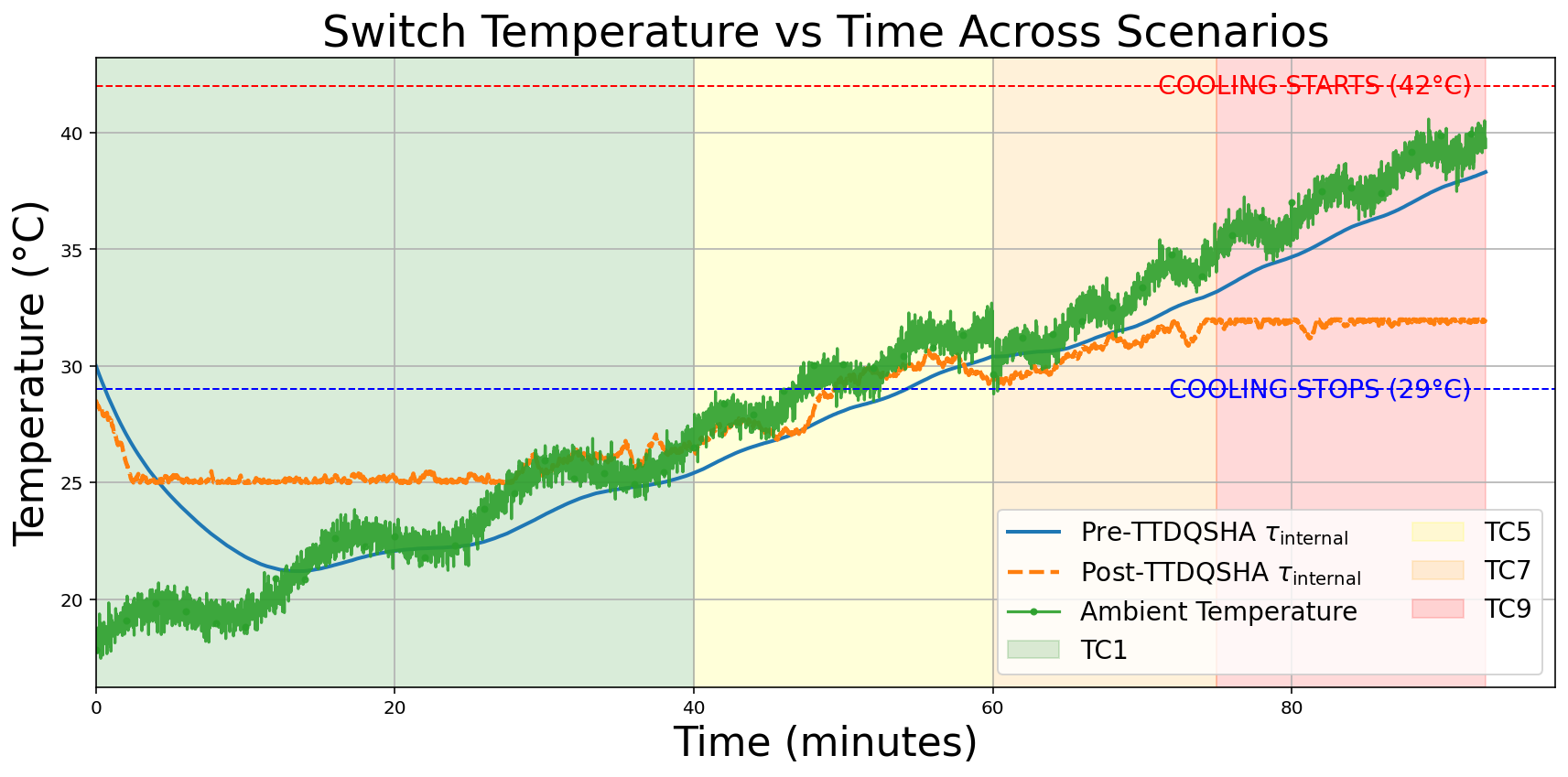}
    \caption{Switch internal temperature dynamics under sequential thermal-stress scenarios}
    \label{fig:temp}
\end{figure}

%Across the nine test cases, the results show that the proposed \texttt{TTDQSHA} framework responds more effectively to thermal stress than the baseline. 
In the scenarios with the highest heat and traffic levels (TC4, TC6, and TC9), the DQN agent quickly detected when switches started to overheat and rerouted traffic to cooler, redundant paths.
This rerouting reduced the load on the affected switches, thereby lowering their internal temperatures because less packet processing generated less heat. Consequently, the reduced load enabled the cooling system to operate more efficiently. As shown in Figure~\ref{fig:temp}, the moving-average temperature curves gradually decline once these actions are taken.
The results demonstrate that the \texttt{TTDQSHA} framework can both identify rising thermal risks and take corrective action to stabilize switch temperatures, even under challenging environmental and operational conditions.

%Reviewer comment: Provide full details on baseline implementations (polling cadence, observability, tuning). Ensure comparisons use equivalent information and polling frequencies. Discuss control overhead vs. benefit tradeoffs.

%The proposed DQN self-healing framework outperformed the conventional baseline by significantly reducing response time, improving latency, minimizing packet loss, and integrating temperature awareness into traffic engineering decisions. 

\subsection{Self-Healing Agent Quantitative Evaluation Against State-of-the-art Benchmarks}
The effectiveness of a self-healing agent hinges on its ability to promptly detect and respond to traffic anomalies and thermal fluctuations in Ethernet switches. 
Table \ref{tab:qeval} summarizes key metrics used to compare the proposed self-healing agent with existing benchmarks. The primary objective across all approaches is to reduce end-to-end latency and link utilization below predefined thresholds. The baseline approach consistently yields the highest latencies and link utilizations (up to 9.35 ms and 78.9\%, respectively). This is due to its static routing decisions and lack of real-time feedback, which make it ill-suited to handle stochastic disruptions effectively, leading to increased packet loss (3.2\%) and retransmissions. Also, the baseline approach exhibited considerably slower reaction times. Topology changes are processed according to the \texttt{OFPT-HELLO} interval of $1$--$9$~s, and rerouting flows may take up to $40$~s, illustrating a significant responsiveness gap relative to \texttt{TTDQSHA}.

\begin{table}
\centering
\caption{SELF-HEALING AGENT QUANTITATIVE EVALUATION AGAINST STATE-OF-THE-ART BENCHMARKS}
\label{tab:qeval}
\resizebox{\columnwidth}{!}{
\begin{tabular}{|l|c|c|c|c|}
\hline
\textbf{Metric}       & \textbf{Baseline} & \textbf{ANFIS} & \textbf{DTPRO} & \textbf{TTDQSHA} \\
\hline
Delay (ms)            & \{$\downarrow$1.6 - 9.35$\uparrow$\}   & \{$\downarrow$0.65 - 2.4$\uparrow$\}  & \{$\downarrow$0.9 - 3.6$\uparrow$\}  & \{$\downarrow$0.75 - 1.9$\uparrow$\}   \\ \hline
Utilization (\%)      & 78.9             & 70.1           & 68.5           & 65.3             \\ \hline
Packet Loss (\%)      & 3.2              & 1.7            & 1.4            & 0.9              \\ \hline
Convergence Time (s)  &  -             & 5.8           & 6.5            & 4.3              \\ \hline
Control Overhead (KB/s) & 2.3             & 10.4           & 18.7           & 12.6             \\ \hline
Decision Time (ms)    & 0.25             & 9.7            & 15.3           & 8.5              \\ \hline
Stability (updates/s) & -            & 0.9            & 1.6            & 0.8              \\ \hline
Retransmissions (×1k) & 13               & 7              & 6              & 4                \\ \hline
Scalability           & Struggles        & Stable         & Degrades       & Stable           \\ \hline
\end{tabular}
}
\end{table}

In contrast, \texttt{ANFIS} employs adaptive fuzzy logic to adjust routing rules based on observed network conditions, thereby improving handling of transient disruptions. This results in significant improvements in latency (reduced to 0.65 ms), utilization (70.1\%), and packet loss (1.7\%). However, its rule-based design imposes moderate control overhead (10.4 KB/s) and slower decision times (9.7 ms), reflecting the computational cost of fuzzy inference and membership function tuning.
\texttt{DTPRO} yields lower utilization (68.5\%) and packet loss (1.4\%) than \texttt{ANFIS}; its higher control overhead (18.7 KB/s) and longer decision times (15.3 ms) indicate higher computational complexity, which limits scalability. Moreover, \texttt{DTPRO}'s performance degrades under higher network loads due to increased retraining demands and longer convergence time (6.5 s).

The proposed \texttt{TTDQSHA} achieves the most favorable trade-offs overall. By combining threshold-triggered actions with DQN, it maintains low latency (0.75–1.9 ms), reduces average link utilization to 65.3\%, and significantly lowers packet loss to 0.9\%. The threshold-triggering mechanism limits unnecessary control traffic, balancing control overhead (12.6 KB/s) with stable update frequency (0.8 updates/s) to enable real-time responsiveness. Faster convergence time (4.3 s) and decision time (8.5 ms) demonstrate computational efficiency and suitability for dynamic environments, while lower retransmissions further confirm robust traffic management. Scalability tests show that \texttt{TTDQSHA} and \texttt{ANFIS} maintain performance under increased network load, whereas \texttt{DTPRO} degrades. 
Overall, these results highlight the control-overhead-benefit trade-off across all agents. While ANFIS and DTPRO incur higher overhead and decision times to achieve their performance gains, \texttt{TTDQSHA} maintains competitive accuracy with substantially lower and more stable control traffic, offering a more favorable balance between responsiveness and operational cost.

%Figure \ref{fig:radial} radial plot further illustrates the distinct performance advantages of the proposed approach over existing benchmarks using normalized values for the different metrics. Overall, these findings highlight the architectural and algorithmic strengths of \texttt{TTDQSHA}} in delivering resilient, high-performance network management under stochastic disruptions. 
    
\subsection{Industrial application recommendations and insights into future work}
%\subsubsection{Response time and convergence}
\noindent
Key industrial application recommendations are: 

\circled{1} Industrial networks must comply with strict Quality of Service (QoS) and Service Level Agreement (SLA) requirements, as defined by standards such as IEC 61400-25 and IEC 61850 for OSS-critical infrastructure. Therefore, the cold start challenges common to learning algorithms necessitate that the agent be thoroughly trained and validated in a controlled environment before deployment in production systems.

\circled{2} The self-healing agent is deployed within a microservices architecture, where different modules interact and may experience latency that can impact the decision-making process of the \texttt{DECIDE} module. Consequently, adequate computational and memory resources must be provisioned, particularly when handling large-scale, heterogeneous data planes. 

\circled{3} Initial delays in processing and the inter-module communication can be mitigated through time-based synchronization, which ensures timely, data-driven decision-making. This, in turn, enhances the agent's responsiveness and readiness for deployment in production environments.
%\subsubsection{Insights into future work}
\\ \\ \noindent
Building upon the limitations and insights of this study, we propose the following avenues for future research: 

\circled{1} This study focused solely on flash events triggered by benign traffic flows over resource-constrained networks, primarily due to legitimate WPP operational activities. 
However, it is plausible that intruders may engineer DDoS attacks that mimic such flash events, especially under extreme operational scenarios such as TC7 through TC9. 
In such cases, \texttt{TTDQSHA} may struggle to accurately distinguish between legitimate and malicious traffic, particularly when both exhibit similar flow characteristics, such as matching \texttt{ETHTypes}. 
As a result, the agent may misclassify malicious traffic as benign, allowing attacks to persist undetected. The SDNC may respond by enforcing traffic rate-limiting policies, potentially leading to considerable packet loss and degraded network performance.
To mitigate this, future work should focus on enhancing the proposed \texttt{TTDQSHA} framework by integrating a dedicated \texttt{DEFENSE} module. This module would independently manage cyber incident responses using anomaly detection and isolation techniques implemented within the \texttt{ORIENT} module, thereby ensuring more robust, reliable self-healing behavior.

\circled{2} Our testbed utilized a reduced-scale model of clustered WPPs, comprising 20 WTGs per WPP, due to computational and memory resource constraints. Future research should investigate the performance and scalability of the self-healing agent in managing larger, more complex network environments, such as clustered WPPs with over 100 WTGs each. Such expanded network configurations would significantly increase the state-space input and demand greater computational and memory resources to empirically validate the operational efficiency of the self-healing framework under these scaled conditions.

%Thermal control was particularly effective in TC4, TC6, and TC9. In these cases, the DQN agent adapted routing paths to offload overheated switches and activated HVAC systems only when strictly necessary. This two-pronged approach enabled efficient temperature regulation with minimal fan use, preserving switch longevity and reducing unnecessary energy consumption. 
%Notably, in constrained scenarios (TC8, TC9), where fewer redundant paths existed, the DQN agent demonstrated robust prioritization strategies, ensuring that time-critical traffic remained unaffected even under thermal and bandwidth pressure.
\section{CONCLUSION} \label{part5}
%%%%%% Problem Statement
Software-defined IIoT-Edge networks in offshore WPPs periodically experience flash events in benign traffic flows, leading to network congestion and intermittent service interruptions. These flash events result from regular offshore WPP operations and maintenance activities such as (i) state changes in wind turbines, which often trigger bursts of data exchange between the centralized protection and control system and the actuators, (ii) SCADA polling during maintenance checks or diagnostic cycles, and (iii) scheduled firmware or software updates.
% Which are undesirable for the offshore WPP operators.
Furthermore, the offshore environment's intricacies expose these IIoT-Edge network equipment to extreme temperature profiles, which can cause physical damage, degrade equipment performance, and reduce their overall lifespan by 50\%.
%%%% Paper objective and Methodology (Algorithm)
To mitigate these two types of stochastic disruptions, this paper designed and implemented an externally adapted \textit{``threshold-triggered Deep Q-Network self-healing framework''} that autonomously monitored the operating temperature profiles of the IIoT-Edge network equipment, detected traffic anomalies that violated QoS and SLAs, and initiated corrective actions.
These corrective actions rerouted traffic to avoid congested paths and overheated switches and throttled low-priority traffic based on service type classification in extreme test case scenarios.
%%%% Results and Discussion
Simulation results demonstrate that the proposed \texttt{TTDQSHA} outperformed the baseline approach (Dijkstra + ECMP) by approximately 53.84\% and the state-of-the-art benchmarks, such as ANFIS self-healing agent by 13.1\% and DTPRO by 21.5\%, when evaluated on a super-spine leaf data-plane switch network. These findings demonstrated the potential of autonomic network architectures to enhance the resilience of software-defined IIoT-Edge networks, particularly in high-availability and consistent-performance-demanding application scenarios.

%%%%%%%%%%%%%%%%%%%%%%%%%%%%%%%%%%%%%%%%%%%%%%%%%%%%%%%%
\section*{ACKNOWLEDGEMENT}
The authors thank the OT Systems Engineering department at Ørsted Wind Power (Gentofte, Denmark, and Warsaw, Poland) for their invaluable input on the network design, and Mikkel Peter Sidoroff Gryning for the acquisition and funding of the Microsoft Azure testbed. The contributions of the authors affiliated with Ørsted to this paper are made in their capacity and do not represent the official views or position of Ørsted.

%%%%%%%%%%%%%%%%%%%%%%%%%%%%%%%%%%%%%%%%%%%%%%%%%%%%%%%%

\bibliography{refs}
\bibliographystyle{ieeetr}

%%%%%%%%%%%%%%%%%%%%%%%%%%%%%%%%%%%%%%%%%%%%%%%%%%%%%%%%

%%%%%%%%%%%%%%%%%%%%%%%%%%%%%%%%%%%%%%%%%%%%%%%%%%%%%%%%%
%\section*{Biography Section}
\vspace{-1cm}
\begin{IEEEbiography}[{\includegraphics[width=1in,height=1.25in,clip,keepaspectratio]{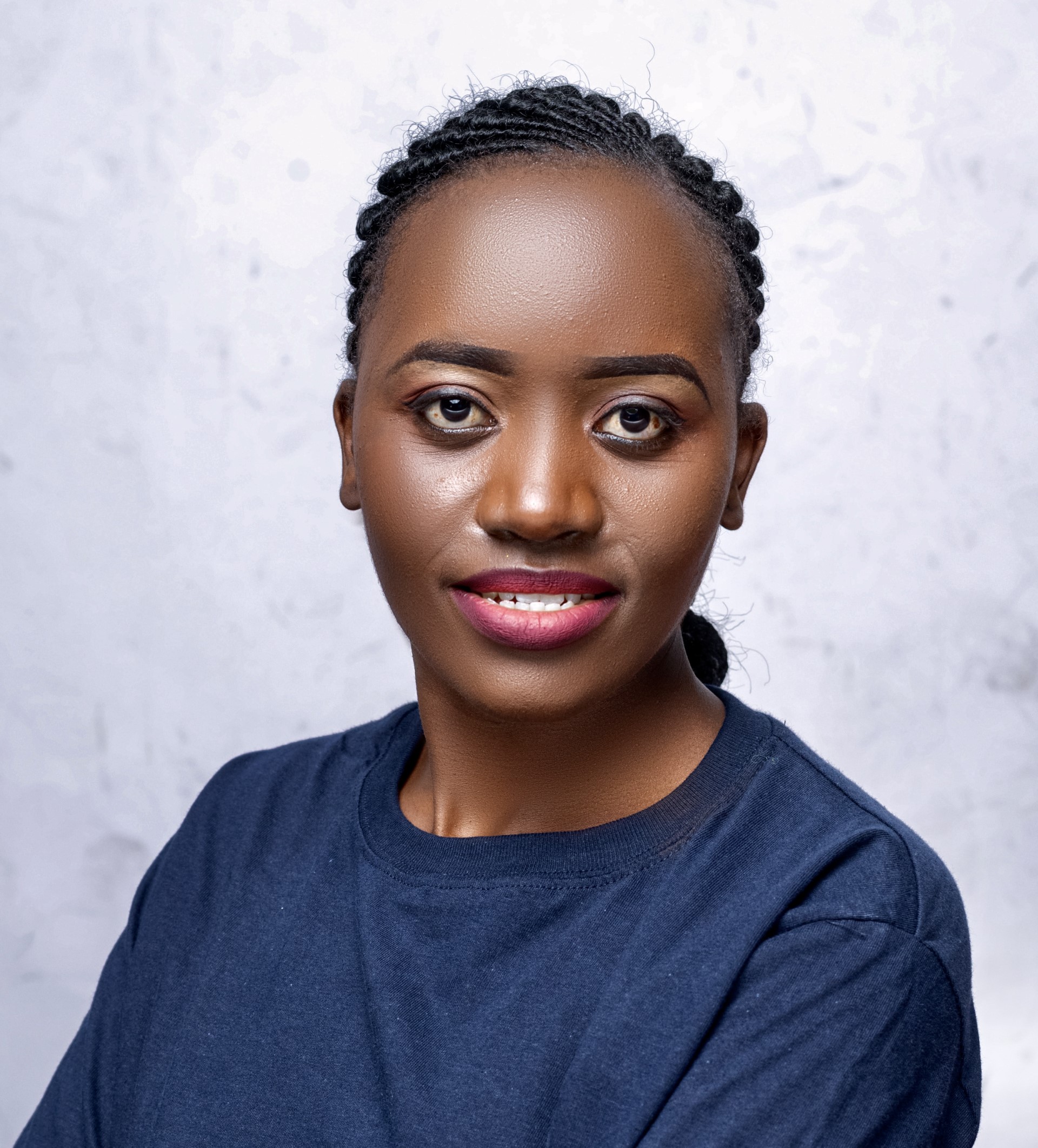}}]{Agrippina Mwangi} Received her BSc. in Electrical and Electronic Engineering from Dedan Kimathi University of Technology and MSc. in Electrical and Computer Engineering from Carnegie Mellon University. 
She is pursuing a Ph.D. in the Energy \& Resources Group of the Copernicus Institute of Sustainable Development, Utrecht University (The Netherlands). 
Her research interests are IoT Edge architectures, SDN/NFV, autonomous networks, Reinforcement Learning, and offshore wind.
\end{IEEEbiography}
\vspace{-1cm}
\begin{IEEEbiography}[{\includegraphics[width=1in,height=1.25in,clip,keepaspectratio]{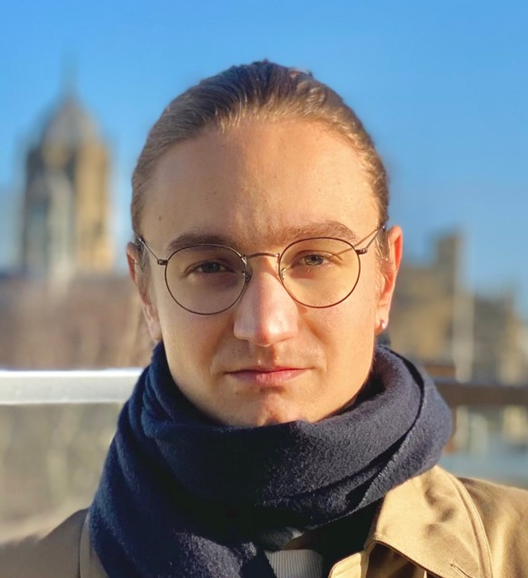}}]{León Navarro-Hilfiker}
Received his MEng in Optimal Control from the University of Oxford and is a Member of InstMC. He is an Operational Technology (OT) Systems Security Engineer, Engineering, Ørsted North America (Providence RI, USA). His research interests include the stability of power systems, optimal distributed control of power grids, cybersecurity, privacy-preserving computations, and communication network control.
\end{IEEEbiography}
\vspace{-1cm}
\begin{IEEEbiography}[{\includegraphics[width=1in,height=1.25in,clip,keepaspectratio]{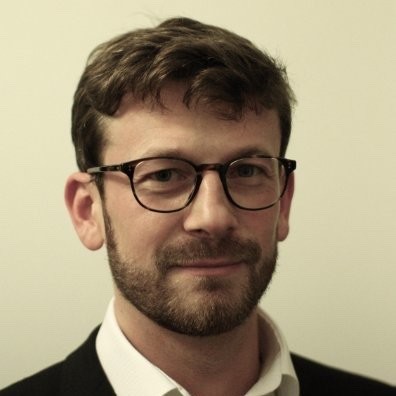}}]{Lukasz Brewka}
Received his M.Sc. and Ph.D. in Electronics and Telecommunication from the Technical University of Denmark. He heads the Operational Technology (OT) Systems Engineering Department, Engineering, Ørsted Wind Power. He is in charge of the design and operation of secure, robust, and resilient communication systems for critical wind farm infrastructure. His research interests include uPnP QoS architecture, network automation and orchestration, and self-healing networks.
\end{IEEEbiography}
\vspace{-1cm}
\begin{IEEEbiography}[{\includegraphics[width=1in,height=1.25in,clip,keepaspectratio]{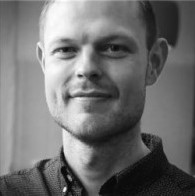}}]{Mikkel Gryning}
Received his
M.Sc. and Ph.D. in Electrical Engineering
from the Technical University of Denmark.
Working as Chief Systems Design Specialist
at Ørsted Wind Power.
His research interests include
power system stability robustness evaluation,
hybrid power plant integration and design, and
communication systems for offshore wind power plants.
\end{IEEEbiography}
\vspace{-1cm}
\begin{IEEEbiography}[{\includegraphics[width=1.1in,height=1.5in,clip,keepaspectratio]{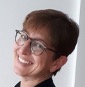}}]{Elena Fumagalli} Received
her MSc. in Nuclear Engineering
from Politecnico di Milano (Italy) and a
Ph.D. in Energy Engineering from Università
di Padova (Italy). She is currently
an associate professor at Utrecht
University, Department of Sustainable Development,
Section Energy and Resources. Her research interests
span electricity market design, network regulation,
and smart grids.
\end{IEEEbiography}
\vspace{-1cm}
\begin{IEEEbiography}[{\includegraphics[width=1in,height=1.25in,clip,keepaspectratio]{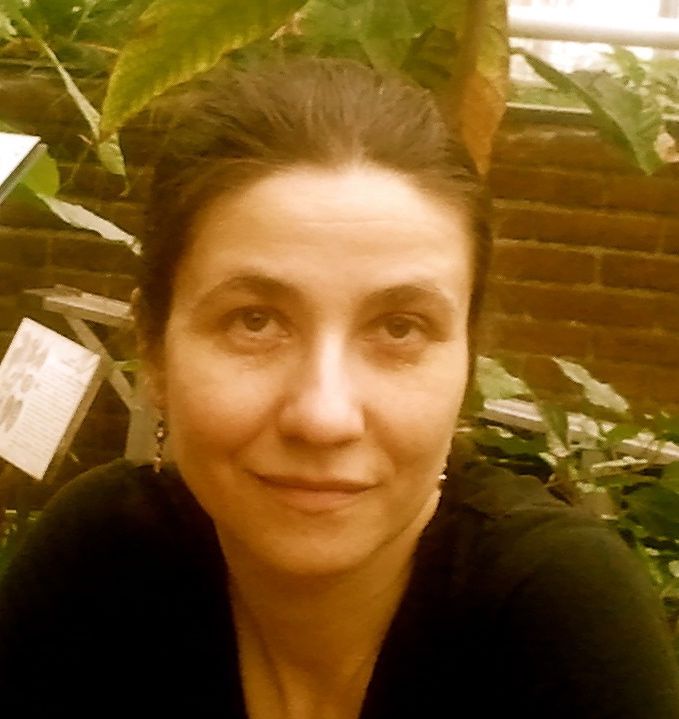}}]{Madeleine Gibescu}
a full professor
in the area of Integration of Intermittent
Renewable Energy in the Copernicus
Institute of Sustainable Development,
Faculty of Geosciences, Utrecht
University. Dr. Gibescu received her
Ph.D. in Electrical Engineering from the
University of Washington, Seattle, Washington, U.S. in
2003. Her research interests are in the modeling and simulation
of power systems and electricity markets with a large
penetration of renewable energy sources. Since 2004, Prof.
Gibescu has worked in the Netherlands in various academic
capacities at TU Delft, TU Eindhoven, and Utrecht University,
leading research in smart grids and system integration
of wind energy and solar photovoltaics.
\end{IEEEbiography}
%%%%%%%%%%%%%%%%%%%%%%%%%%%%%%%%%%%%%%%%%%%%%%%%%%%%%%%%%%

\end{document}